\documentclass[%
 reprint,
 amsmath,amssymb,
 aps,
]{revtex4-2}

\usepackage{graphicx}
\usepackage{dcolumn}
\usepackage{bm}
\usepackage{hyperref}
\usepackage{slashed}
\usepackage{subfloat}
\usepackage{pgfplots,subfigure}
\usepackage{orcidlink}
\usepackage{geometry}
\definecolor{acsblue}{RGB}{17,76,139}

\begin{document}
\fontsize{7.8}{8.8}\selectfont
\preprint{APS/123-QED}

\title{Quantum confinement of scalar bosons in the Bonnor-Melvin spacetime: uniform magnetic field and rainbow gravity effects}

\author{Omar Mustafa}
\email{omar.mustafa@emu.edu.tr (Corr. Author)}
\affiliation{Department of Physics, Eastern Mediterranean University, 99628, G. Magusa, north Cyprus, Mersin 10 - Türkiye}

\author{Abdullah Guvendi}
\email{abdullah.guvendi@erzurum.edu.tr}
\affiliation{Department of Basic Sciences, Erzurum Technical University, 25050, Erzurum, Türkiye}

\date{\today}

\begin{abstract}
\setlength{\parindent}{0pt}

{\fontsize{7.8}{8.8}\selectfont 
We present an exact analytical study of Klein-Gordon (KG) scalar bosons and antibosons confined in the Bonnor-Melvin (BM) spacetime under a uniform magnetic field, incorporating rainbow gravity (RG) corrections with a positive cosmological constant. The cosmological constant partitions spacetime into an infinite sequence of confinement domains bounded by impenetrable barriers. Within the first allowed domain, the KG equation reduces to a hypergeometric differential equation, yielding closed-form expressions for both the energy spectra and the radial wavefunctions in terms of hypergeometric polynomials. Two representative RG models, inspired by the Magueijo-Smolin framework and loop quantum gravity (LQG), produce Planck-scale bounded, symmetric particle-antiparticle spectra. A distinctive feature of the curved magnetized geometry is the collapse of all magnetic quantum states $m \neq 0$ onto the $m = 0$ level for each radial excitation, a degeneracy absent in flat spacetime. Increasing the cosmological constant partially lifts this collapse, establishing a direct link between the global spacetime curvature and the local quantum structure. Radial probability density analysis further shows that stronger magnetic fields enhance spatial localization, confining bosons into static or rotating ring-like configurations with nodal architectures that evolve systematically with quantum numbers. These findings reveal how gravitational confinement, topology, magnetic fields, and Planck-scale corrections jointly govern the spectral and spatial properties of relativistic quantum fields in curved and magnetized backgrounds.} 

\end{abstract}


\maketitle


\section{Introduction}

\vspace{0.10cm}
\setlength{\parindent}{0pt}

Einstein’s theory of general relativity (GR) \cite{1.1} describes gravity as a manifestation of the curvature of spacetime. The attempt to unify GR with quantum mechanics, often referred to as quantum gravity (QG), offers fundamental insights into the behavior of quantum systems in curved backgrounds. This connection has motivated many studies examining how spacetime curvature influences quantum spectroscopic properties, especially for scalar and spinor particles described by the KG and Dirac equations, respectively \cite{1.2}. The present study focuses on KG particles in such settings.

\vspace{0.10cm}
\setlength{\parindent}{0pt}

Magnetic fields are central to many astrophysical and cosmological phenomena, including stellar structures, accretion disks, galactic centers, magnetars \cite{1.3,1.4}, and heavy-ion collisions \cite{1.5,1.6,1.7}. The simultaneous presence of strong gravitational and magnetic fields near compact objects has led to considerable interest in general relativistic models that incorporate both effects. {One prominent example is the BM universe, an exact solution of the Einstein-Maxwell equations that describes a static spacetime with cylindrical symmetry, sourced by a magnetic field aligned with the symmetry axis and self-consistently embedded in its own gravitational field \cite{R1.1,R1.2,R1.3,R1.4,R1.4.1}}. In this model, the magnetic field contributes to the energy-momentum tensor and thereby affects the geometry of spacetime. A positive cosmological constant $\Lambda > 0$ is introduced to stabilize the spacetime and preserve its symmetry. The BM spacetime metric is given in \cite{R1.1} as
\begin{equation}
ds^{2} = -dt^{2} + dr^{2} + \tilde{\sigma}^{2} \sin^{2} \left( \sqrt{2\Lambda} r \right) d\varphi^{2} + dz^{2},  
\label{I.1}
\end{equation}
where \(z\in[-\infty,\infty],\,{\varphi\in[0,2\pi]},\,r\in[0,\infty]\), and \(\Lambda\) in units of inverse length squared. Following the coordinate transformation \(\sqrt{2\Lambda}r \rightarrow \rho\) and \(\tilde{\sigma}\sqrt{2\Lambda}\varphi \rightarrow \phi\), as in {\v{Z}}ofka’s treatment \cite{R1.1}, the line element \eqref{I.1} becomes
\begin{equation}
    ds^{2} = -dt^{2} +\frac{1}{2\Lambda} \left(d\rho^{2} + \sin^{2} \left(\rho \right)\, d\phi^{2}\right) + dz^{2},  
\label{I.1.1}
\end{equation}
where \(H = \frac{1}{\sqrt{2}}\, \sin (\rho)\) denotes the magnetic field \cite{R1.1}. 

\vspace{0.10cm}
\setlength{\parindent}{0pt}

Topological defects such as domain walls are expected in various unified field theories \cite{R1.5,R1.6,R1.7,R1.8}, and their effects on quantum dynamics have been examined in \cite{R1.9,R1.10,R1.11,R1.12,R1.13,R1.14,R1.15,R1.16,R1.17,R1.18,AH,A-epjc,AO-advanced,SAO,AO,R1.19,R1.20,R1.21}. The relationship between gravitational geometry and quantum systems remains a key motivation for studying such configurations, and the BM spacetime provides a suitable model for this purpose \cite{R1.1,R1.2,R1.3,R1.22,R1.23,AO-2,AAS}.

\vspace{0.10cm}
\setlength{\parindent}{0pt}

In the high-energy (ultraviolet) regime, the energy of a test particle can influence the structure of spacetime, leading to a modification of the standard dispersion relation \cite{R1.24,R1.25,R1.26,R1.27,R1.28,R1.29,R1.30}:
\begin{equation}
E^{2}f_{_0}\left(u\right) ^{2}-p^{2}f_{_1}(u)
^{2}=m_\circ^{2}. \label{I.2}
\end{equation}
This form appears in various QG theories, such as string theory, LQG, and non-commutative geometry \cite{R1.31,R1.32,R1.33}, as well as in some recent research framed within RG, including: energy-momentum distributions of a cylindrical black hole in rainbow gravity \cite{R1.33.1}; generalized uncertainty principle corrections in Rastall-Rainbow Casimir wormholes \cite{R1.33.2}; fermions moving around strong gravitational sources \cite{R1.33.3}; and the exploration of the physical properties of the strange star SAX J1808.4--3658 in rainbow gravity \cite{R1.33.4}.

\textcolor{red}{Consequently}, the BM metric \eqref{I.1.1} under RG becomes
\begin{equation}
ds^{2}=-\frac{dt^{2}}{f_{_0}\left( u\right) ^{2}}+\frac{1}{f_{_1}\left( u\right) ^{2}}\left(\frac{1}{2\Lambda} \left[d\rho^{2}+\sin ^{2}\left( \rho\right)
\,d\phi ^{2}\right]+dz^{2}\right),  \label{I.3}
\end{equation}
where \(f_{_i}(u);\;i=0,1,\) are the so-called rainbow functions that obey the infrared limit \(\lim\limits_{u\rightarrow 0}f_{_i}\left( u\right) =1\), and \(u=|E|/E_{p}\) satisfies \(0\leq u \leq 1\). This ensures a consistent transition to the classical regime and an equal treatment of particles and antiparticles \cite{R1.30,R1.34,R1.35,R1.35.1,R1.35.2}. The functions must preserve \(E_p\), the Planck energy, as an invariant scale. We refer to the metric in (\ref{I.3}) as the BM-RG spacetime. In our recent studies \cite{R1.35.1,R1.35.2}, we showed that the BM spacetime contains singularities at \(\rho=\sqrt{2\Lambda }r=\tau \,\pi\), with \(\tau=0,1,2,\cdots\). These singularities correspond to infinite potential barriers, acting effectively as impenetrable walls. The surfaces at \(r = \frac{\tau \,\pi}{\sqrt{2\Lambda}}\) do not constitute domain walls in the strict field-theoretic sense, since a domain wall normally separates distinct vacuum phases. Instead, they arise as geometric or topological defects due to degeneration of the angular metric component \(g_{\phi\phi}\). At these radial locations, the angular component degenerates, producing singularities in the spacetime structure. Notably, this array of impenetrable surfaces imposes effective confinement for fields within each radial domain, resembling domain-wall-like behavior phenomenologically, while their origin is purely geometric rather than associated with any phase transition of the field. The confinement mechanism in the BM spacetime is therefore emphasized. As a result, quantum particles are confined between two successive infinite walls, such as the region \(\rho =\sqrt{2\Lambda}r\in[0, \pi]\). This confinement agrees with {\v{Z}}ofka's result \cite{R1.1}, confirming the presence of a boundary at \(\sqrt{2\Lambda }r=\pi\), and makes any treatment that assumes \(r\rightarrow \infty\) ($\rho\to\infty$)  invalid \cite{R1.4.1,R1.35.3}. In \cite{R1.35.1}, we analyzed KG particles in a nonuniform magnetic field in the BM-RG background and showed that the resulting problem is conditionally exactly solvable using the general Heun function \(H_G(a, q, \alpha, \beta, \gamma, \delta, z)\).

\vspace{0.10cm}
\setlength{\parindent}{0pt}

In the present study, we consider KG particles in a uniform external magnetic field in the BM-RG background. In Section \ref{sec:2}, we formulate the KG equation in this setting and show that it reduces to a hypergeometric differential equation. This leads to exact solutions in terms of hypergeometric polynomials of order \(n\). Unlike the case of a nonuniform magnetic field in \cite{R1.35.1}, where we were only able to find a conditionally exact solution using general Heun polynomials. We derive the energy spectrum and wave functions by analyzing the associated hypergeometric series. Here, our analyses interestingly show that, contrary to standard quantum mechanics, where one would expect a magnetic field to at least partially lift the degeneracy associated with the magnetic quantum number \(m = \pm|m|\), we observe no such effect. Moreover, we find that as the magnetic field strength increases, all \(m\)-states appear to collapse onto the \(m = 0\) state for a given radial quantum number \(n\). Nevertheless, the cosmological constant \(\Lambda\) is observed to lift the degeneracy associated with \(m\), even within its standard range \(0 < \Lambda \ll 1\). In our recent study \cite{R1.35.1} of KG particles in a nonuniform external magnetic field within the BM RG framework, such surprising degeneracy behaviors were not observed. In Section \ref{sec:3}, we examine the effect of RG using two sets of rainbow functions. The first set, proposed by Magueijo and Smolin, is given by \(f_{_0}(u) = 1/(1 - \tilde{\epsilon}|E|)\) and \(f_{_1}(u) = 1\), with \(\tilde{\epsilon} = \epsilon/E_p\), motivated by varying speed of light models \cite{R1.36}. The second set, inspired by LQG \cite{R1.38,R1.39}, uses \(f_{_0}(u) = 1\) and \(f_{_1}(u) = \sqrt{1 - \tilde{\epsilon}|E|^\nu}\), with \(\nu = 1,2\). Both sets comply with the principles of RG and maintain \(E_p\) as an upper energy bound \cite{R1.34,R1.35,R1.35.1,R1.35.2,R1.40}. Our concluding remarks are presented in Section \ref{sec:4}.

\section{KG-particles in a uniform magnetic field in BM-RG-spacetime background}\label{sec:2}

\vspace{0.10cm}
\setlength{\parindent}{0pt}

In this section, we formulate a non-perturbative wave equation by incorporating the effects of the RG framework. For the BM-RG spacetime, the non-zero components of the contravariant metric tensor \(g^{\mu\nu}\) associated with the line element (\ref{I.3}) are given by
\begin{equation}
    g^{00}=-f_{_0}(u)^2,\,\,g^{11}=2\Lambda f_{_1}(u)^2,\,\,g^{22}=\frac{2\Lambda f_{_1}(u)^2}{\sin(\rho)^2},\,\,g^{33}=f_{_1}(u)^2. \label{II.2}
\end{equation}
The determinant of the covariant metric tensor is
\[
\text{det}(g_{\mu\nu})=g=-\frac{ \sin(\rho)^2}{ 4\Lambda^2 f_{_1}(u)^6 f_{_0}(u)^2}.
\]
We proceed by considering a KG particle and antiparticle in the presence of an external magnetic field within the BM background. The governing equation reads
\begin{equation}
    \frac{1}{\sqrt{-g}}\left(\partial_\mu -ieA_\mu\right)\sqrt{-g}\,g^{\mu\nu} \,\left(\partial_\nu-ieA_\nu \right)\, \psi(t,\rho,\phi,z)=m_\circ ^2\, \psi(t,\rho,\phi,z),\label{II.3}
\end{equation}
where \(m_\circ\) denotes the rest mass energy (or rest energy) of the particle. We consider the electromagnetic four-vector potential in the form \(A_\mu = (0, 0, A_{\phi}, 0)\), with
\begin{equation*}
A_{\phi}=-\frac{ \mathcal{B}_\circ}{2\Lambda} \cos(\rho)\Rightarrow 
\textbf{B}=\mathcal{B}_\circ\,f_{_1}(u)^2 \, \hat{z},
\end{equation*}
as obtained from the electromagnetic field invariant \(\mathcal{F}_{\mu\nu}\mathcal{F}^{\mu\nu}/2
=||\textbf{B}||^2-||\textbf{E}||^2/{c^2}\). It is evident that in the absence of RG corrections, where \(f_{_1}(u)=1\), the magnetic field is strictly uniform. However, the presence of RG effects modifies the field, rendering it energy-dependent. To determine the influence of the BM spacetime on KG bosons under a uniform magnetic field, we employ the ansatz \(\psi(t,\rho,\phi,z)=e^{i\,(m\,\phi+k\,z-E\,t)}\, R(\rho)\) in Eq.~(\ref{II.3}), which yields the following radial differential equation:
\begin{equation}
R ^{\prime \prime }\left( \rho\right) +\frac{1}{\tan  \rho }R
^{\prime }\left( \rho\right) +\left( \mathcal{E} -\frac{\left[{m}+\tilde{\mathcal{B}}\,\cos\rho\right]^2}{\sin^2 
\rho}\right) R\left( \rho\right) =0,  \label{II.4}
\end{equation}
where
\begin{equation}
 \tilde{\mathcal{B}}=\frac{e\mathcal{B}_{\circ}}{{2\Lambda}},\quad\mathcal{E} =\frac{f_{_0}\left( u\right) ^{2}E^{2}-m_{\circ}^{2}}{ 2\Lambda f_{_1}\left( u\right) ^{2}}-\frac{k^{2}}{2\Lambda},\label{II.5}
\end{equation}
and the integer \(m=0,\pm1,\pm2,\cdots\) corresponds to the magnetic quantum number. This expression can be further simplified into the form
\begin{equation}
    R ^{\prime \prime }\left( \rho\right) +\frac{1}{\tan  \rho }R
^{\prime }\left( \rho\right) +\left( \tilde{\mathcal{E}} -\frac{\eta^2}{\sin^2\rho}-\frac{2\,{m}\,\tilde{\mathcal{B}}\,\cos\rho}{\sin^2\rho}\right) R\left( \rho\right) =0, \label{II.6}
\end{equation}
where we have introduced \(\tilde{\mathcal{E}}=\mathcal{E}+\tilde{\mathcal{B}}^2\) and \( \eta^2=({m}^2+\tilde{\mathcal{B}}^2) \Rightarrow \eta=\left|\sqrt{({m}^2+\tilde{\mathcal{B}}^2)}\right|\). To recast this equation into a one-dimensional Schrödinger-like form, which facilitates the identification of the effective gravitational potential, we define \(R(\rho)=U(\rho)/\sqrt{\sin(\rho)}\). Substituting into Eq.~(\ref{II.6}) leads to
\begin{equation}
    U''(\rho)+\left[\tilde{\lambda}-\frac{\left(\eta^2-1/4\right)}{\sin^2\rho}-\frac{2\,{m}\,\tilde{\mathcal{B}}\,\cos\rho}{\sin^2\rho}\right]U(\rho)=0, \label{II.7}
\end{equation}
with \(\tilde{\lambda}=\tilde{\mathcal{E}}+1/4\). Accordingly, the effective gravitational potential can be identified as
\begin{equation}
V_{eff}(\rho)=\frac{\left(\eta^2-1/4\right)}{\sin^2\rho}+\frac{2\,{m}\,\tilde{\mathcal{B}}\,\cos\rho}{\sin^2\rho}. \label{II.7.1}
\end{equation}
It should be noted that, due to the small but non-zero cosmological constant \(0<\Lambda\ll1\), and the condition \(\tilde{\mathcal{B}}\neq0\), the effective potential does not impose restrictions on the values of the magnetic quantum number \(m\). Instead, it introduces potential barriers at \(\rho=0,\pi,2\pi,\dots\), confining the quantum particle/antiparticle within the interval \(\rho\in[0,\pi]\). This would suggest the boundary conditions \(U(0)=0=U(\pi)\) for the radial wave function. In the subsequent section, we will derive the exact solution to the corresponding one-dimensional Schrödinger-type equation given in (\ref{II.7}).

\subsection{Geometric Confinement of Klein-Gordon Particles}  \label{sec:2A}

\vspace{0.10cm}
\setlength{\parindent}{0pt}

We begin by analyzing the KG test particles in a uniform magnetic field within BM spacetime induced confinement and described by the metric in (\ref{II.7}), using the substitutions \(x=\cos\rho \to \rho=\arccos(x)\), and
\begin{equation}
    U(x)=(x-1)^{\alpha}\,(x+1)^{\beta}\,F(x), \label{II.2.1}
\end{equation}
where \[\alpha=\frac{1}{4}\left(\sqrt{4\,\eta^2+4\,\gamma+1}+1\right),\quad \beta=\frac{1}{4}\left(\sqrt{4\,\eta^2-4\,\gamma+1}+1\right),\] \(\gamma=2\,{m}\,\tilde{\mathcal{B}}\), and \(\eta^2={m}^2+\tilde{\mathcal{B}}^2\). Note that for \(x=1\) (\(\rho=0\)) and \(x=-1\) (\(\rho=\pi\)), the function \(U(x)=0\), analogous to the standard textbook infinite potential well, represented by the first two impenetrable walls in the BM spacetime scenario. Next, we apply the change of variables \(y=(x+1)/2\), leading to
\begin{equation}
   (y^2-y)F''(x)+\left[(a+b+1)y-c\right]F'(y)+ab\,F(y)=0,\label{II.2.2}
\end{equation}
where
\begin{gather}
    a=\alpha+\beta+\sqrt{\tilde{\lambda}},\quad b=\alpha+\beta-\sqrt{\tilde{\lambda}},\quad c=2\beta+\frac{1}{2}. \label{II.2.3}
\end{gather}
It is evident that equation (\ref{II.2.2}) is the hypergeometric differential equation, which admits the solution \(F(y)= \,_2F_1(a,b,c,y)\), where \(a=-n\) and/or \(b=-n\), would ensure that the hypergeometric power series truncates to a polynomial of order \(n \geq 0\). Furthermore, note that \(\alpha, \beta \geq 1/2\), as can be seen from
\begin{equation}
    \alpha=\frac{1}{4}\left(\sqrt{4({m}+\tilde{\mathcal{B}})^2+1}+1\right),\quad \beta=\frac{1}{4}\left(\sqrt{4({m}-\tilde{\mathcal{B}})^2+1}+1\right).\label{II.2.4}
\end{equation}
It is important to recognize that the term \(\alpha+\beta\) in \(a\) and \(b\) of (\ref{II.2.3}) remains unchanged for \({m}=\pm|{m}|\), where \(\alpha|_{\pm|{m}|}=\beta|_{\mp|{m}|}\). To verify the uniqueness of \(F(y)= \,_2F_1(a,b,c,y)\) as a solution of (\ref{II.2.2}), we consider a power series expansion of the form
\begin{equation}
    F(y)=\sum_{j=0}^{\infty} C_j \,y^{j + \sigma}, \label{II.2.5}
\end{equation}
which yields
\begin{equation}
\begin{split}
&\sum_{j=0}^{\infty} C_j \,\left[ab+(j+\sigma)(j+\sigma+a+b)\right]\,y^{j + \sigma}\\
&= \sum_{j=0}^{\infty} C_{j+1}\left[(j+\sigma+1)(j+\sigma+c)\right]\,y^{j + \sigma}+C_0 \,\sigma\left[\sigma+c-1\right]\,y^{\sigma-1}. \label{II.2.6}
\end{split}
\end{equation}
Since \(C_0\neq0\), it follows that \(\sigma\left[\sigma+c-1\right]=0\). Consequently, \(\sigma=0\) or \(\sigma=-c+1=-2\beta+1/2\). The latter case would cause \(F(y)\to\infty\) as \(y\to0\) (i.e., \(\rho=\pi\in[0,\pi]\)), and must therefore be excluded. We thus adopt \(\sigma=0\), leading to the recurrence relation
\begin{equation}
 C_{j+1}\left[(j+1)(j+c)\right] =C_j \,\left[ab+j(j+a+b)\right]. \label{II.2.7}
\end{equation}
To truncate the power series into a polynomial of degree \(n \geq 0\), we impose the condition \(C_{n+1}=0\) while ensuring \(C_n\neq0\). This requirement yields
\[
ab+n(n+a+b)=0 \Rightarrow a=-n \,\, \text{and/or}\,\, b=-n.
\]
As a result, we obtain \(\tilde{\lambda}=(n+\alpha+\beta)^2\). Consequently, the energy expression becomes
\begin{equation}
\begin{split}
&f_0(u)^2 E^2 =  f_1(u)^2 \mathcal{G}_{nm} + m_{\circ}^2\,; \\ 
&\mathcal{G}_{nm} = 2\Lambda\left[(n+\alpha+\beta)^2-\tilde{\mathcal{B}}^2-\frac{1}{4}\right] + k^2. \label{II.2.9}
\end{split}
\end{equation}
We observe that the invariance of \(\alpha+\beta\) under the transformation \({m} = \pm|{m}|\) is reflected in the spectroscopic structure of KG particles and antiparticles in (\ref{II.2.9}). Specifically, the spectrum exhibits degeneracy with respect to the magnetic quantum number \(m = \pm |m|\), arising from the identity \(\alpha|_{\pm|{m}|} = \beta|_{\mp|{m}|}\). Furthermore, since \(F(y)= \,_2F_1(a,b,c,y)=  \sum_{j=0}^{n} C_j \,y^j\), the full solution reads
\begin{equation}
\begin{split}
U_{nm}(x)&\sim(x-1)^\alpha\,(x+1)^\beta \,\sum_{j=0}^{n} C_j \,\left(\frac{x+1}{2}\right)^j \Rightarrow \\
U_{nm}(\rho)&=\mathcal{N}\,(\cos(\rho)-1)^\alpha\,(\cos(\rho)+1)^\beta \,\sum_{j=0}^{n} C_j \,\left(\cos\left(\frac{\rho}{2}\right)\right)^{2j} \\ 
R_{nm}(\rho)&=\mathcal{N}\,(\cos(\rho)-1)^{\alpha-1/4}\,(\cos(\rho)+1)^{\beta-1/4} \,\sum_{j=0}^{n} C_j \,\left(\cos\left(\frac{\rho}{2}\right)\right)^{2j}.\label{II.2.10}
\end{split}
\end{equation}
Here, the coefficients \(C_j\) are given by the recurrence relation in (\ref{II.2.7}), with \(\alpha\) and \(\beta\) defined in (\ref{II.2.4}). This result will be used in the following analysis to examine the effects of RG corrections on the spectroscopic structure of KG particles and antiparticles in BM spacetime subjected to a uniform external magnetic field. It is clear that both radial functions \(R(\rho)\) and \(U(\rho)\) vanish at the surfaces \(\rho = 0\) and \(\rho = \pi\), in agreement with the standard behavior of quantum wave functions for particles confined within an infinite potential well of width \(\pi\).

\section{RAINBOW GRAVITY EFFECTS}\label{sec:3}

\begin{figure*}[ht!]
\centering
\includegraphics[width=0.2\textwidth]{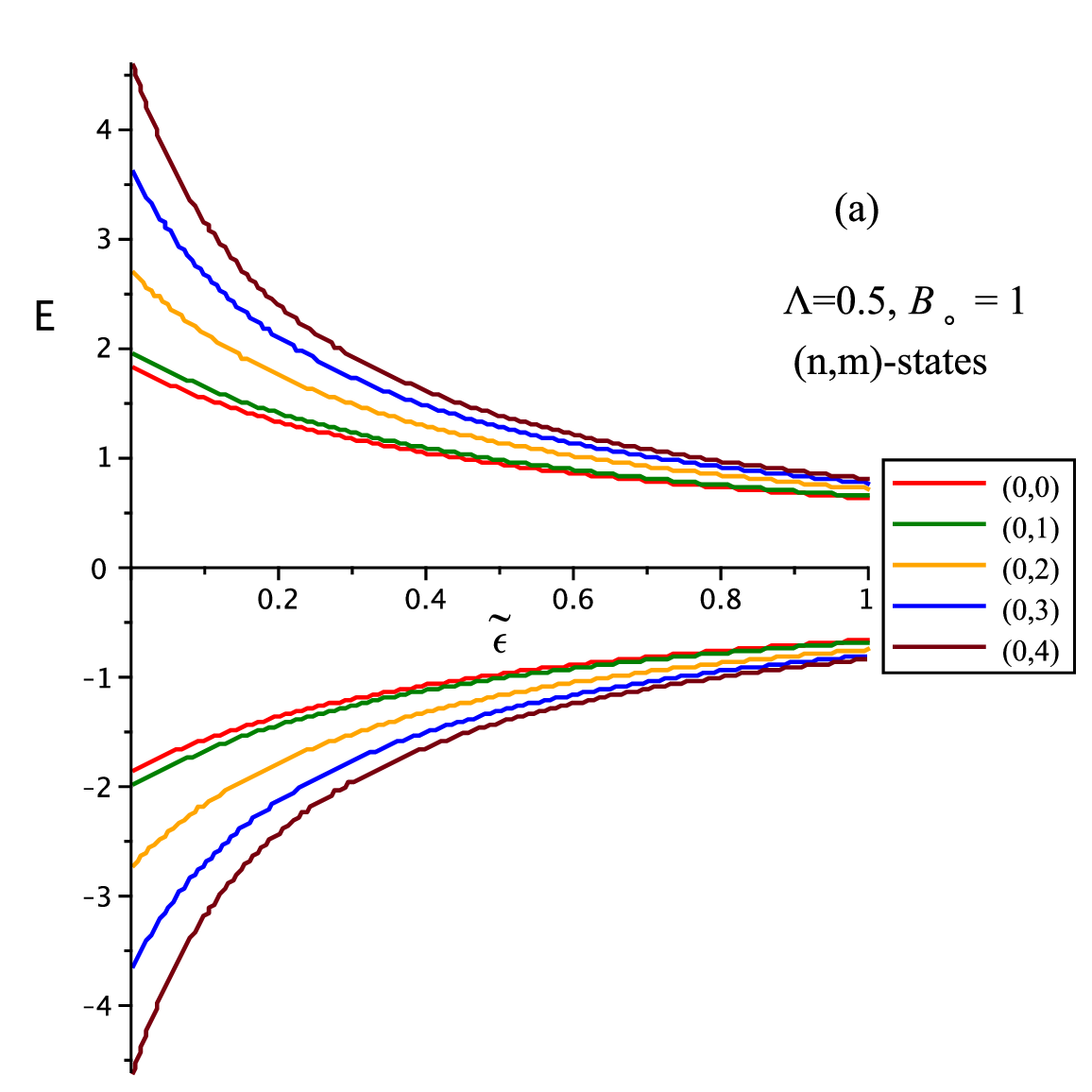}
\includegraphics[width=0.2\textwidth]{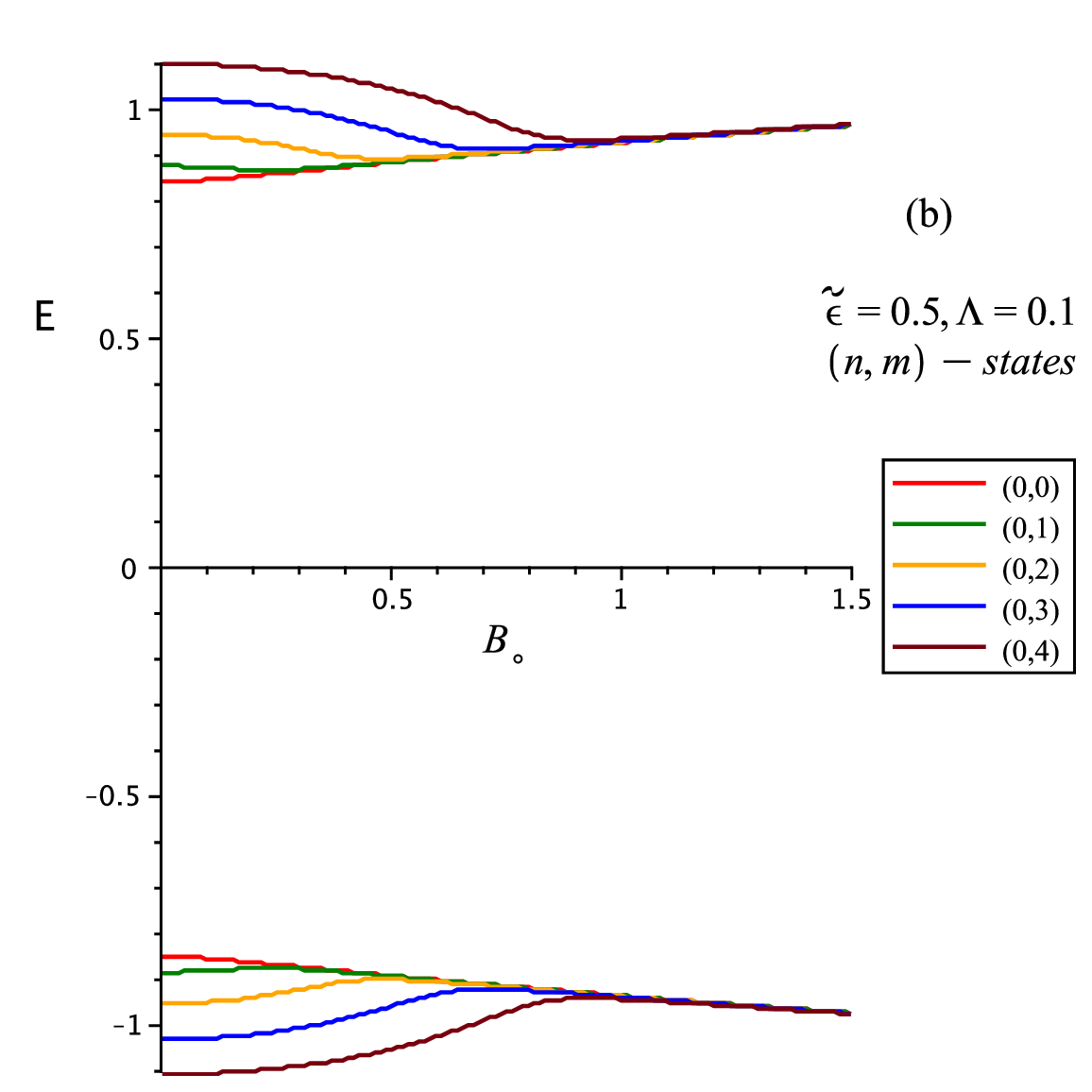}
\includegraphics[width=0.2\textwidth]{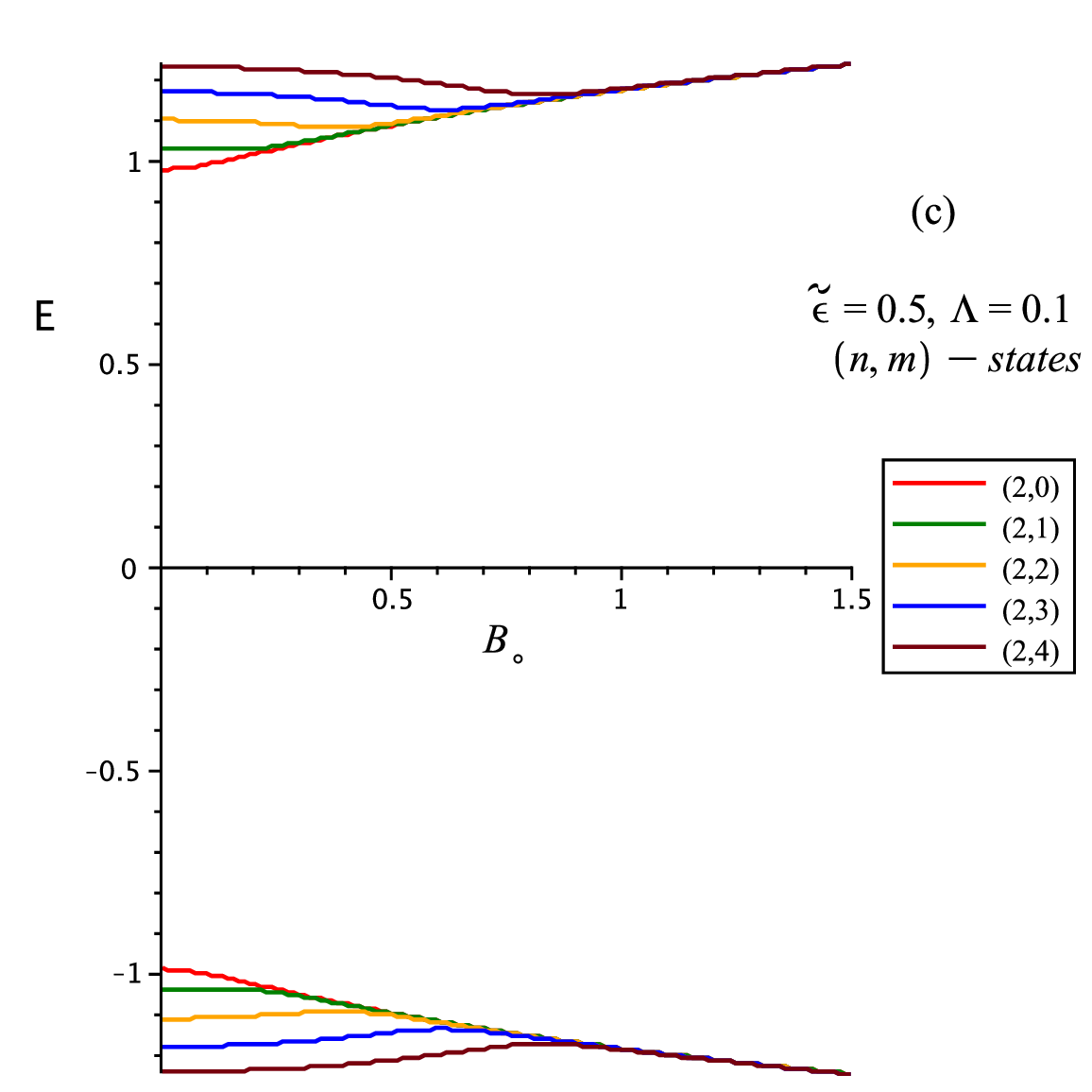}
\includegraphics[width=0.2\textwidth]{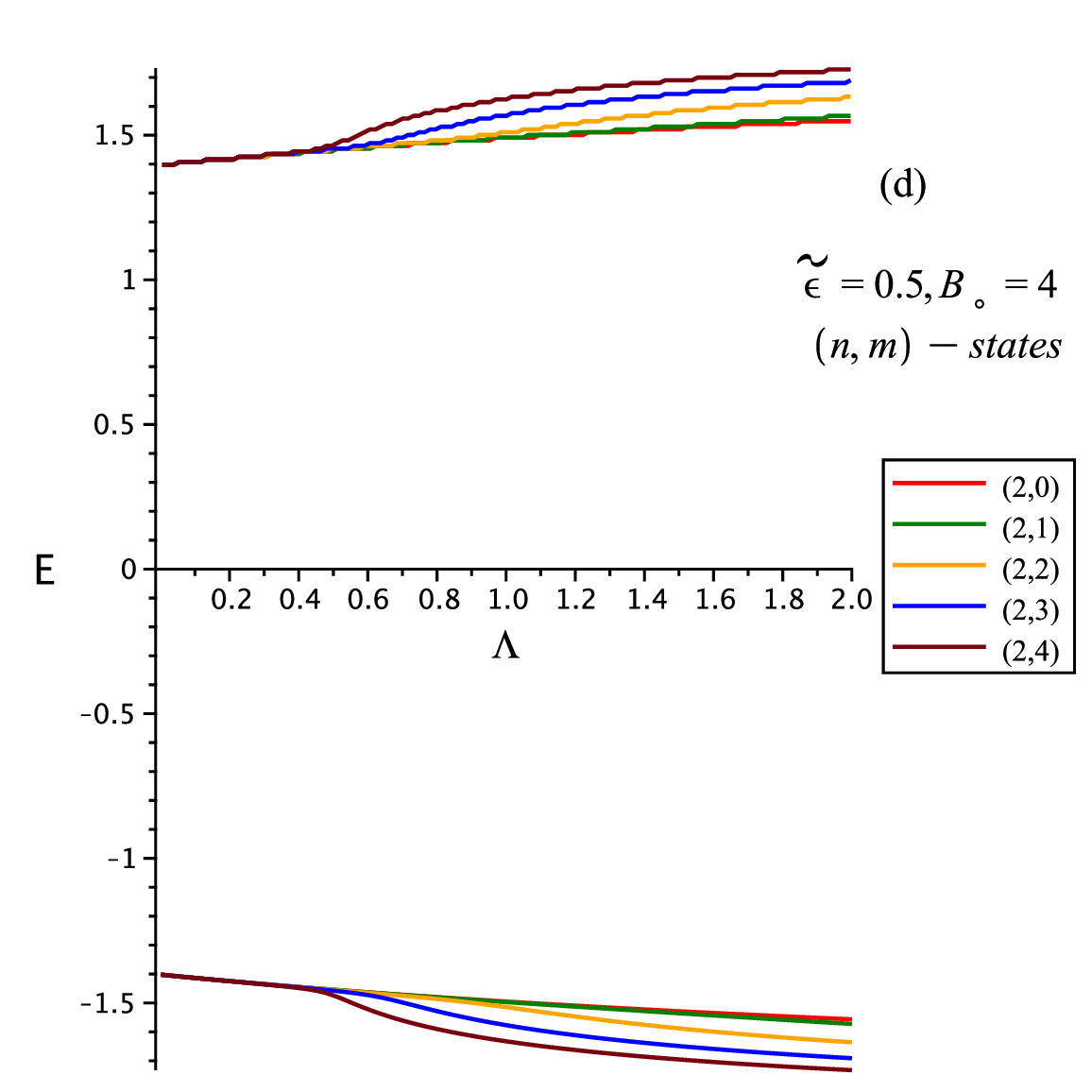}
\includegraphics[width=0.2\textwidth]{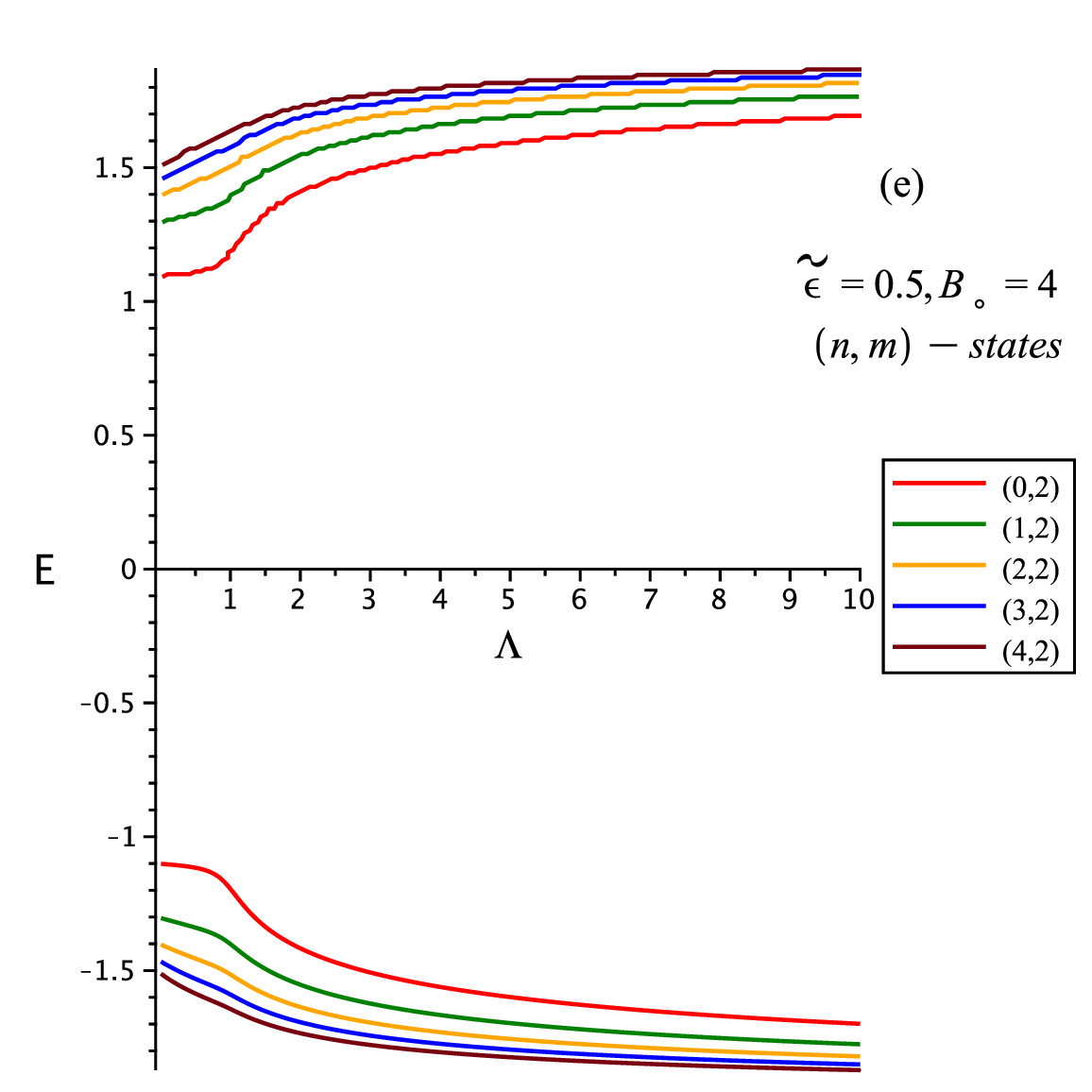}
\includegraphics[width=0.2\textwidth]{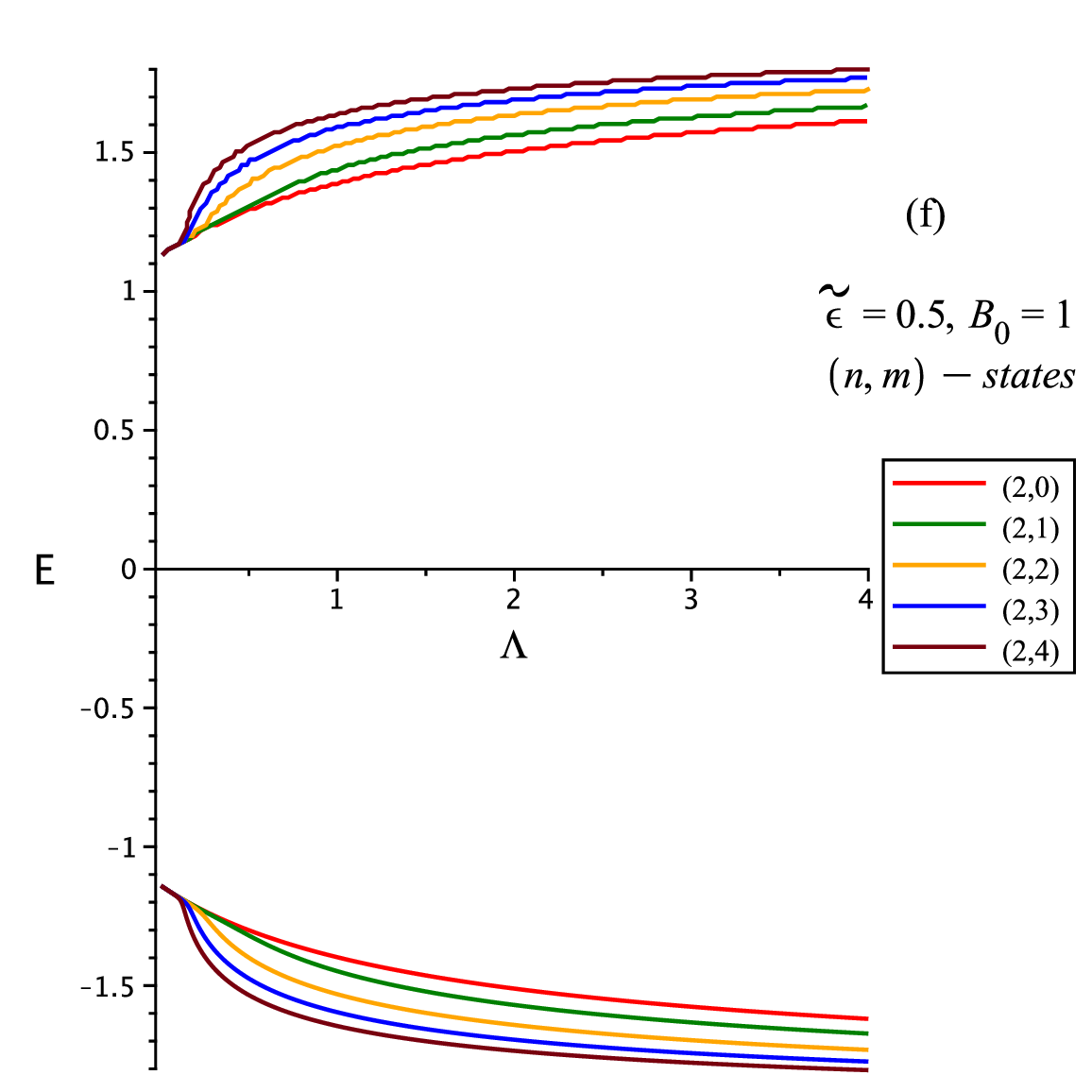}
\caption{\fontsize{7.8}{8.8}\selectfont  The figure displays the energy levels of KG particles and antiparticles given by (\ref{III.1.1}). Specifically, we plot: (a) \(E\) versus \(\tilde{\epsilon}\) for \(n=0\), \(m=0,1,2,3,4\), \(\Lambda=0.5\), and \(\mathcal{B}_\circ=1=e\); (b) \(E\) versus \(\mathcal{B}_\circ\) for \(n=0\), \(m=0,1,2,3,4\), \(\tilde{\epsilon}=0.5\), and \(\Lambda=0.1\); (c) \(E\) versus \(\mathcal{B}_\circ\) for \(n=2\), \(m=0,1,2,3,4\), \(\tilde{\epsilon}=0.5\), and \(\Lambda=0.1\); (d) \(E\) versus the cosmological constant \(\Lambda\) for \(n=2\), \(m=0,1,2,3,4\), \(\tilde{\epsilon}=0.5\), and \(\mathcal{B}_\circ=4\); (e) \(E\) versus \(\Lambda\) for \(n=0,1,2,3,4\), \(m=2\), \(\tilde{\epsilon}=0.5\), and \(\mathcal{B}_\circ=4\); and (f) \(E\) versus \(\Lambda\) for \(n=2\), \(m=0,1,2,3,4\), \(\tilde{\epsilon}=0.5\), and \(\mathcal{B}_\circ=1\).}
\label{fig1}
\end{figure*}

\vspace{0.10cm}
\setlength{\parindent}{0pt}

In this section, we investigate the influence of RG on KG particles and antiparticles by employing two distinct classes of rainbow functions. The first class, originally proposed within the framework of the varying speed of light hypothesis \cite{R1.36}, is defined by
\[
f_0(u) = \frac{1}{1 - \epsilon u} = \frac{1}{1 - \tilde{\epsilon} |E|}, \quad f_1(u) = 1,
\]
where $u = |E|/E_p$ denotes a dimensionless energy scale normalized by the Planck energy $E_p$, and $\tilde{\epsilon} = \epsilon/E_p$ is the corresponding dimensionless rainbow parameter. The second class, inspired by the considerations of LQG \cite{R1.38,R1.39}, is given by
\[
f_0(u) = 1, \quad f_1(u) = \sqrt{1 - \tilde{\epsilon} |E|^{\nu}}, \quad \nu = 1, 2,
\]
with the admissible energy range limited to $0 \leq u = |E|/E_p \leq 1$. These functional forms introduce energy-dependent deformations to the spacetime geometry, encapsulating potential quantum gravitational corrections at trans-Planckian scales.

\begin{figure*}[ht!]
\centering
\includegraphics[width=0.2\textwidth]{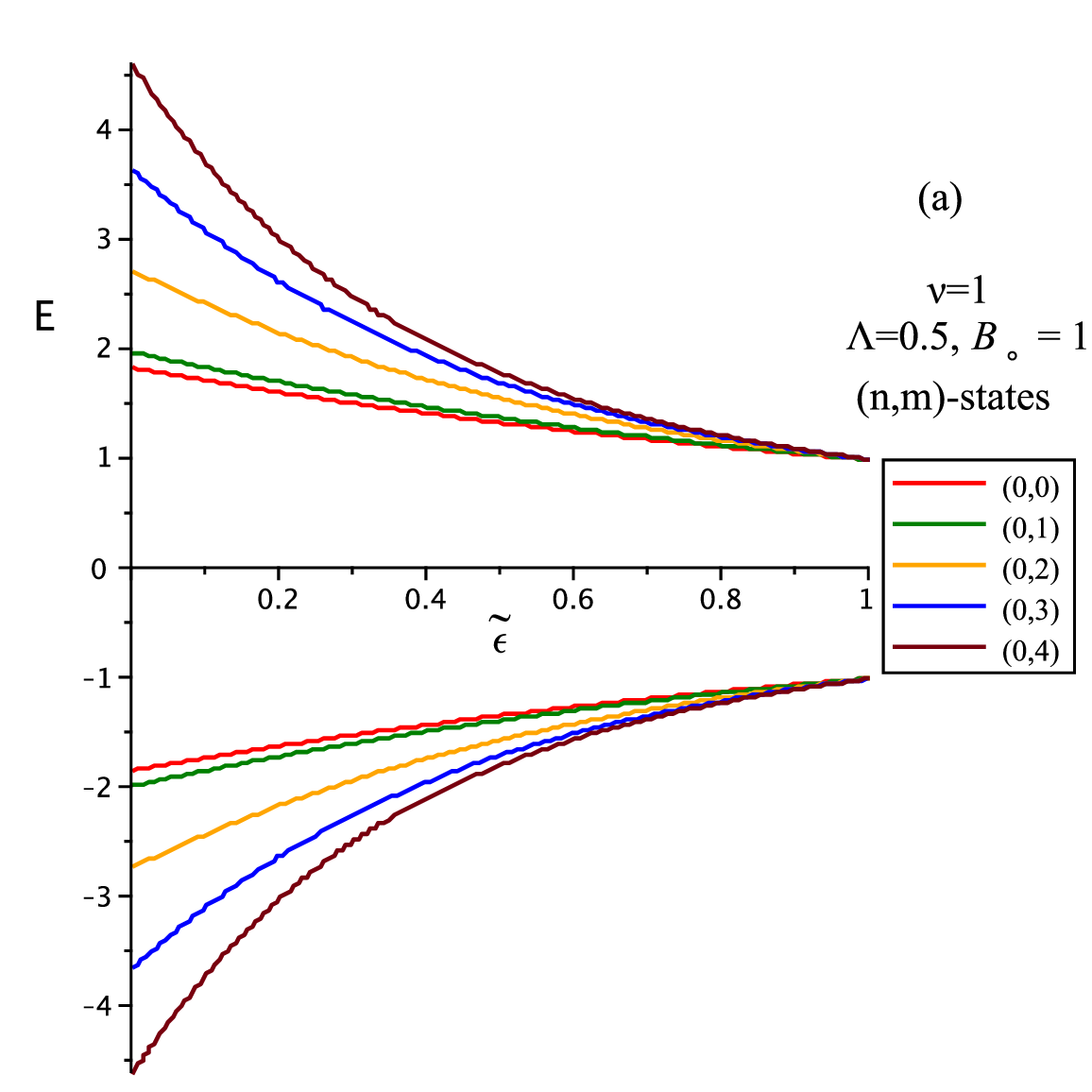}
\includegraphics[width=0.2\textwidth]{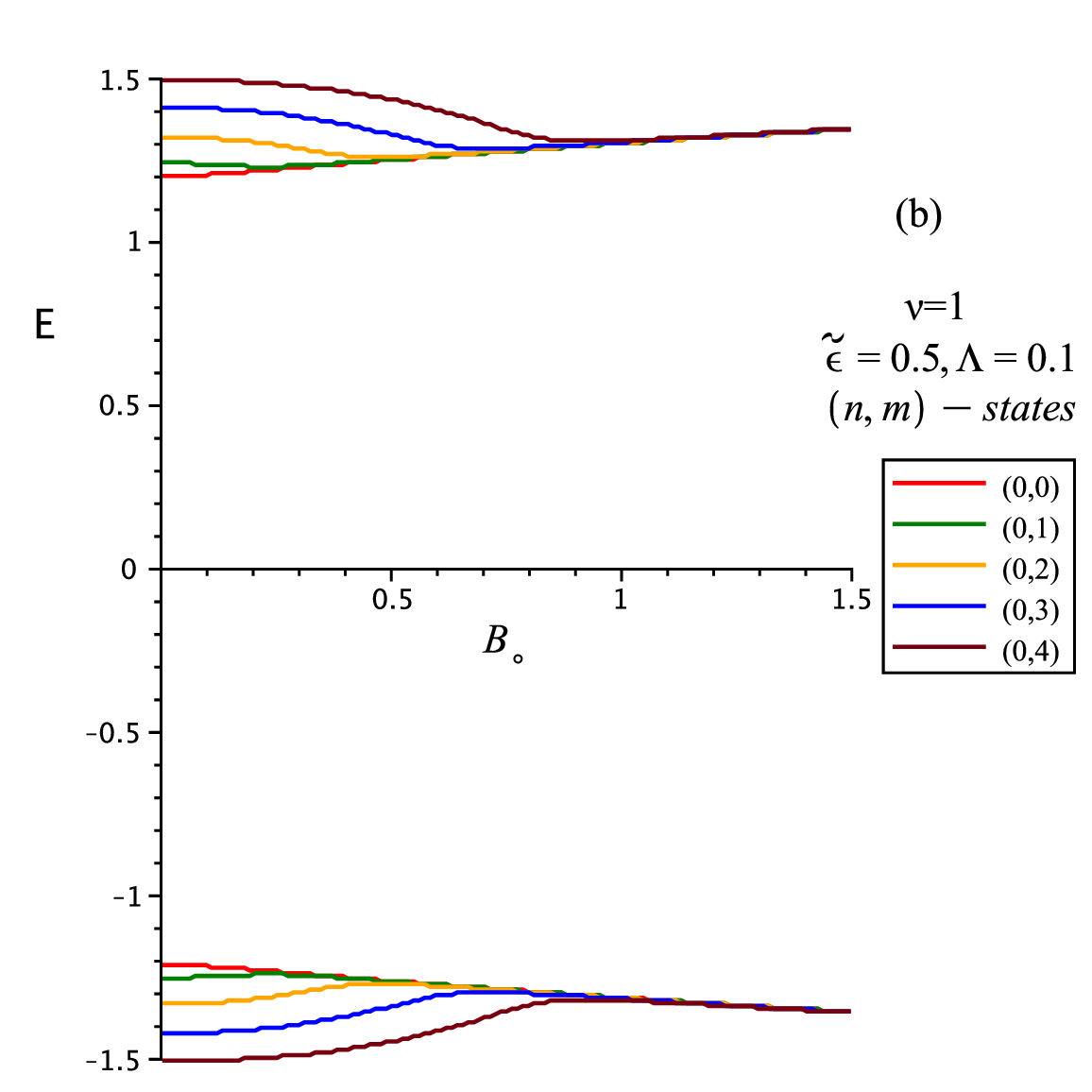}
\includegraphics[width=0.2\textwidth]{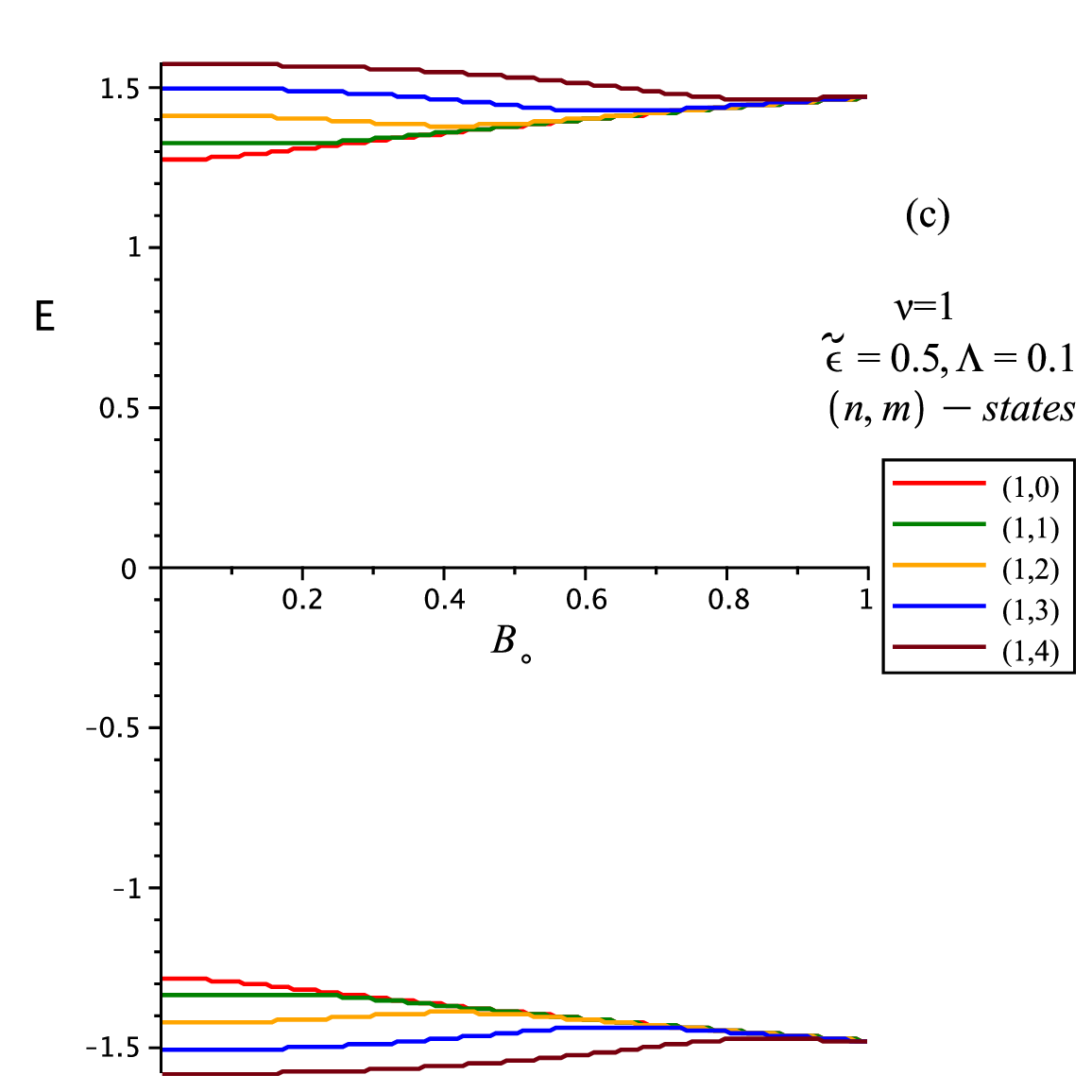}
\includegraphics[width=0.2\textwidth]{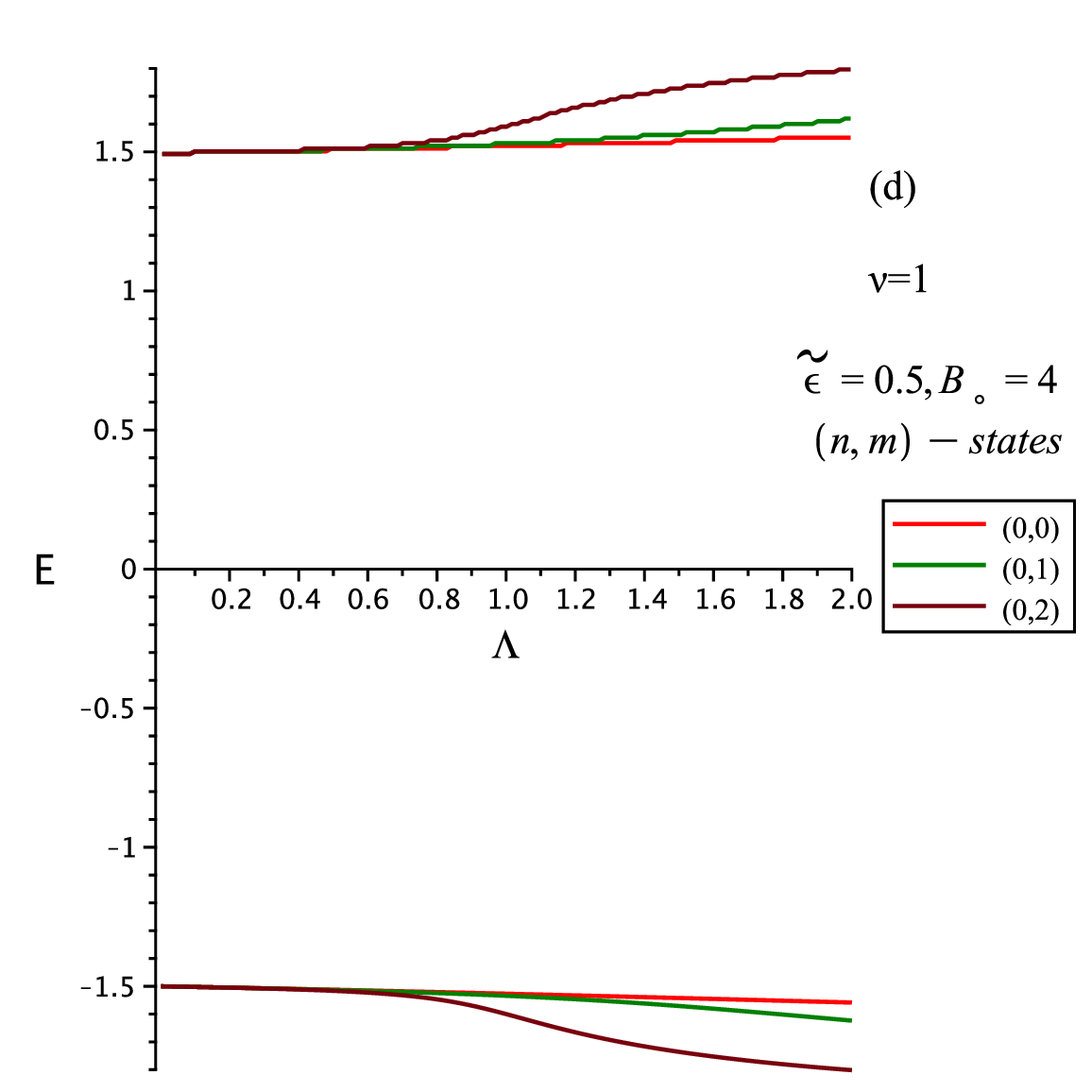}
\caption{\fontsize{7.8}{8.8}\selectfont  The figure illustrates the energy levels of KG particles and antiparticles as given by Eq.~(\ref{III.3.1}). Specifically, the plots show:  (a) $E$ as a function of $\tilde{\epsilon}$ for $n = 0$, $m = 0, 1, 2, 3, 4$, $\Lambda = 0.5$, and $\mathcal{B}_\circ = 1 = e$; (b) $E$ as a function of $\mathcal{B}_\circ$ for $n = 0$, $m = 0, 1, 2, 3, 4$, $\Lambda = 0.1$, and $\tilde{\epsilon} = 0.5$; (c) $E$ as a function of $\mathcal{B}_\circ$ for $n = 1$, $m = 0, 1, 2, 3, 4$, $\Lambda = 0.1$, and $\tilde{\epsilon} = 0.5$;  (d) $E$ as a function of the cosmological constant $\Lambda$ for $n = 0$, $m = 0, 1, 2$, $\mathcal{B}_\circ = 4$, and $\tilde{\epsilon} = 0.5$.
}
\label{fig2}
\end{figure*}

\vspace{0.10cm}

\subsection{Magueijo-Smolin Rainbow Functions}

\setlength{\parindent}{0pt}

This set of rainbow functions, \( f_{_0}\left(u \right) =1/\left( 1-\tilde{\epsilon}\left\vert E\right\vert \right) \), \( f_{_1}\left( u \right) =1 \) (with  $\tilde{\epsilon}=\epsilon/E_p$), is motivated by the hypothesis of varying the speed of light \cite{R1.36}. Substituting this set into the result (\ref{II.2.9}) leads to
\begin{equation}
E^{2} = \tilde{\mathcal{G}}_{nm}\left( 1 - \tilde{\epsilon}\left\vert E \right\vert \right)^{2};\; \tilde{\mathcal{G}}_{nm} = \mathcal{G}_{nm} + m_{\circ}^{2}.
\label{III.1.1}
\end{equation}
It is important to recall that \(\left\vert E \right\vert = E_{+}\) is used for test particles, while \(\left\vert E \right\vert = -E_{-}\) is used for antiparticles. This distinction leads to two separate equations: 
\begin{equation}
E_{+}^{2}\left( 1 - \tilde{\epsilon}^{2} \tilde{\mathcal{G}}_{nm}\right) + 2 \tilde{\epsilon} \, \tilde{\mathcal{G}}_{nm}\,E_{+} - \tilde{\mathcal{G}}_{nm} = 0,
\label{III.1.2}
\end{equation}
for particles, and 
\begin{equation}
E_{-}^{2} \left( 1 - \tilde{\epsilon}^{2} \tilde{\mathcal{G}}_{nm} \right) - 2 \tilde{\epsilon} \, \tilde{\mathcal{G}}_{nm} E_{-} - \tilde{\mathcal{G}}_{nm} = 0,
\label{III.1.3}
\end{equation}
for antiparticles. These can be combined to yield the general result:
\begin{equation}
E_{n,m} = \pm \frac{\sqrt{\tilde{\mathcal{G}}_{nm}}}{1 + \tilde{\epsilon} \sqrt{\tilde{\mathcal{G}}_{nm}}}.
\label{III.1.4}
\end{equation}
This result suggests a clear symmetrization of the KG particle and antiparticle energies with respect to \(E = 0\), as also observed in Figure \ref{fig1}. However, it should be noted that the asymptotic behavior of \(\left\vert E_{\pm} \right\vert\) as \(\mathcal{B}_\circ \to \infty\) approaches a limiting value \(\left\vert E_{\pm} \right\vert \sim 2 = 1/\tilde{\epsilon} = E_{p}/\epsilon\), as seen in Figs. \ref{fig1}(b) and \ref{fig1}(c). Consequently, it follows that \(\left\vert E_{\pm} \right\vert \leq E_{p} \Longrightarrow \left\vert E_{\pm} \right\vert_{\max} = E_{p}\) when \(\epsilon = 1\). Although it is known that the cosmological constant typically satisfies $0<\Lambda\ll1$ ( i.e., \(\Lambda=1.4657\times10^{-52}m^{-2}=3.827\times 10^{-123}l_p^{-2}\), where \(l_p\) is the Planck length), we have extended this range and considered larger values of $\Lambda$ purely for hypothetical exploration. Although it is hypothetical, it allows us to probe theoretical possibilities and emphasizes that the Planck energy \(E_p\) is the maximum possible energy for particles and antiparticles (a natural consequence of the RG proposal) and is yet another invariant, alongside the speed of light, as clearly illustrated in Figs. \ref{fig1}(d), \ref{fig1}(e), and \ref{fig1}(f). According to standard quantum mechanics, one would expect that a magnetic field lifts the degeneracy associated with the magnetic quantum number \(m = \pm|m|\), at least partially. Interestingly, contrary to this expectation, we observe no such effect. In fact, as the magnetic field strength increases, all \(m\)-states appear to collapse into the \(m = 0\) state for a given \(n\). However, the cosmological constant \(\Lambda\) lifts the degeneracy associated with \(m\), even within its standard range \(0<\Lambda<1\), as clearly shown in Figs. \ref{fig1}(d), \ref{fig1}(e) and \ref{fig1}(f). We attribute this to the uniform nature of the magnetic field employed here. In our recent study \cite{R1.35.1} of KG particles in a non-uniform external magnetic field within the BM RG framework, such surprising degeneracy behaviors were not observed.

\begin{figure*}[ht!]
\centering
\includegraphics[width=0.2\textwidth]{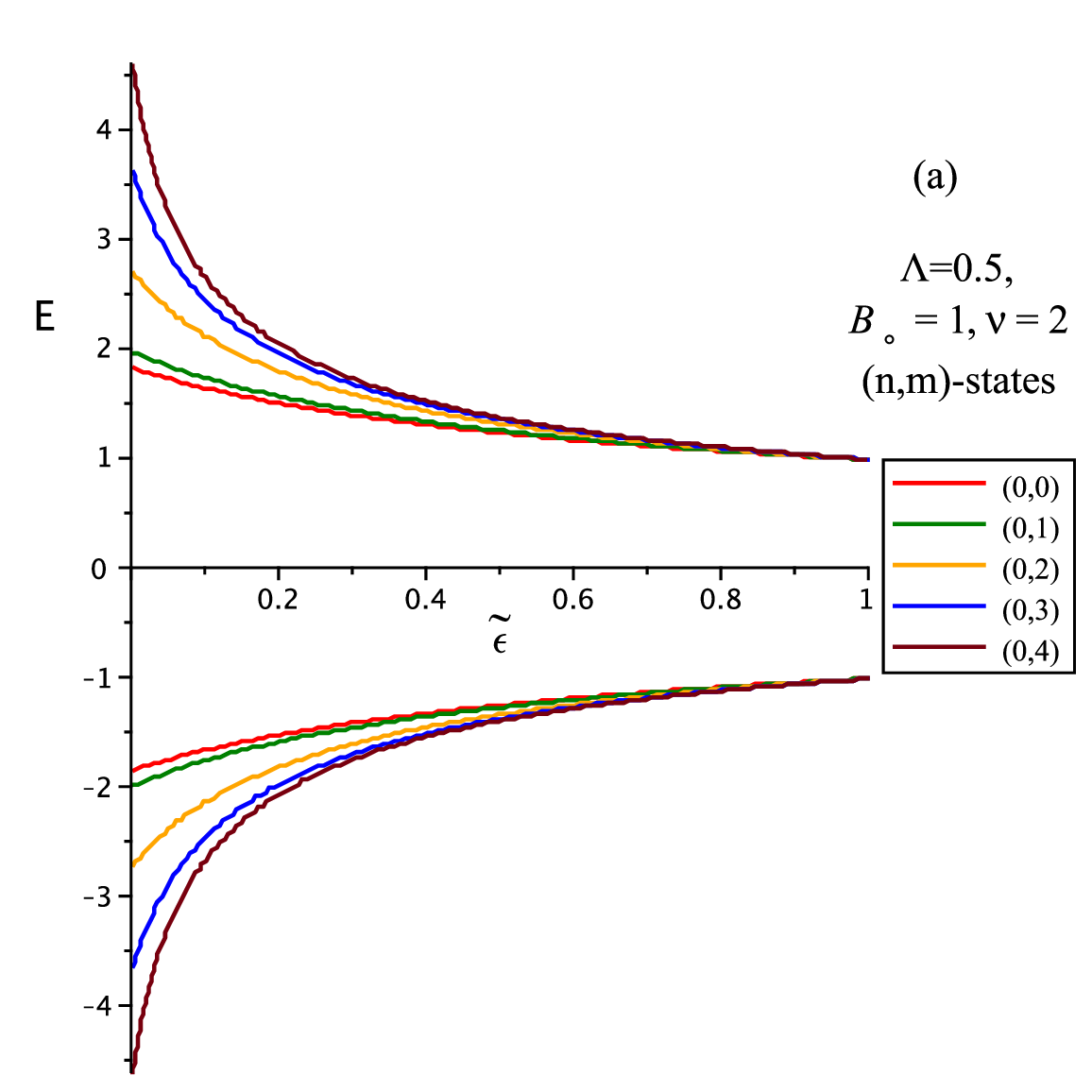}
\includegraphics[width=0.2\textwidth]{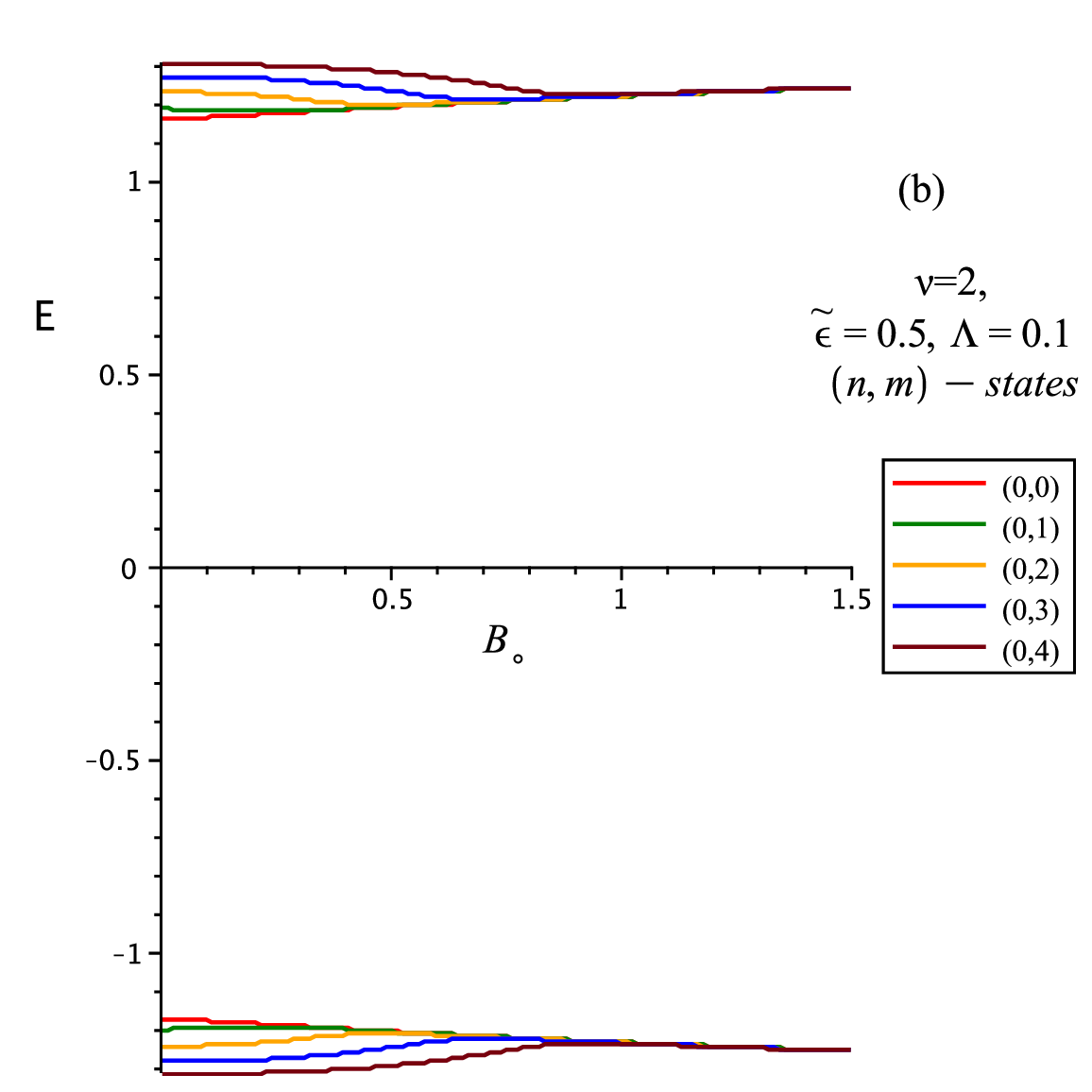}
\includegraphics[width=0.2\textwidth]{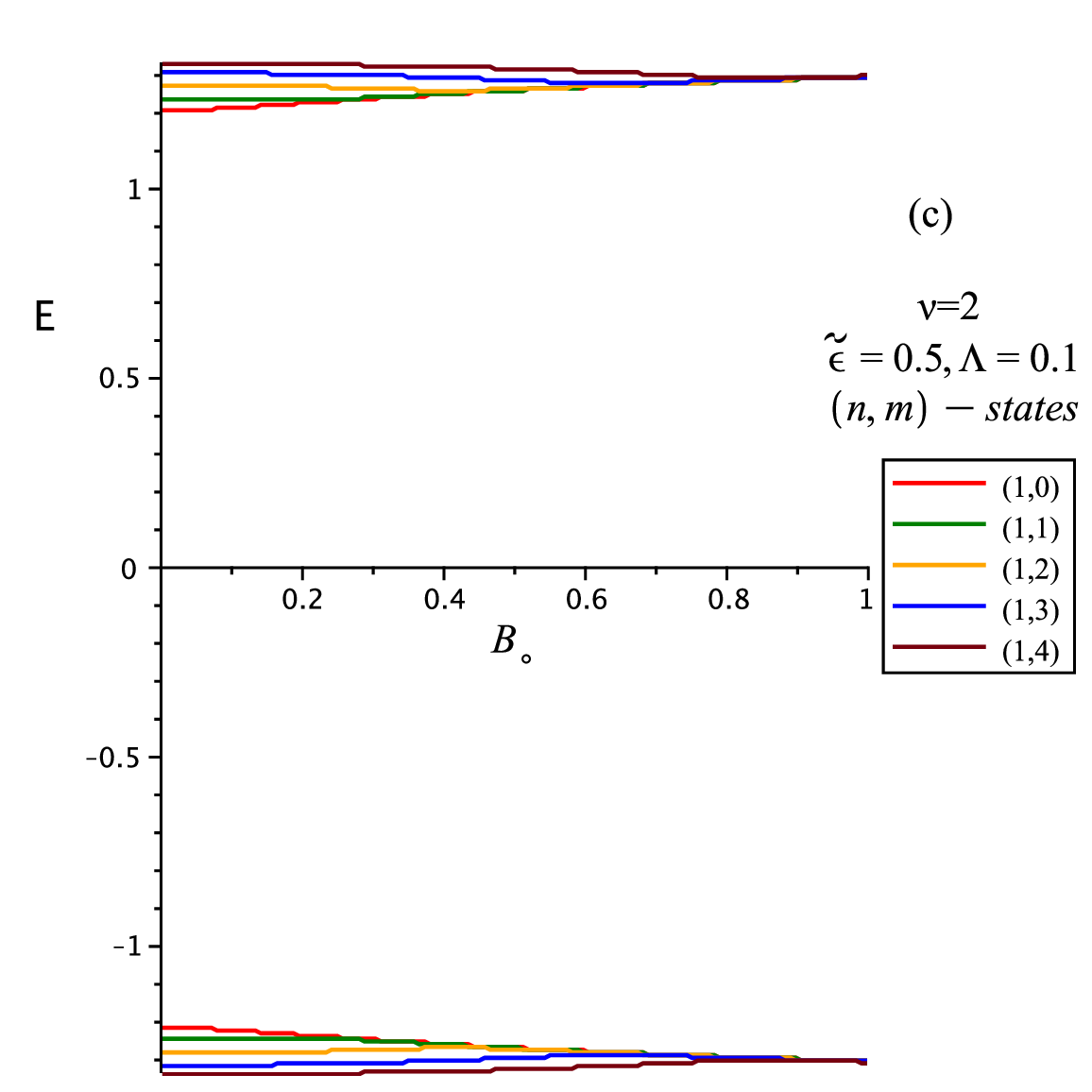}
\includegraphics[width=0.2\textwidth]{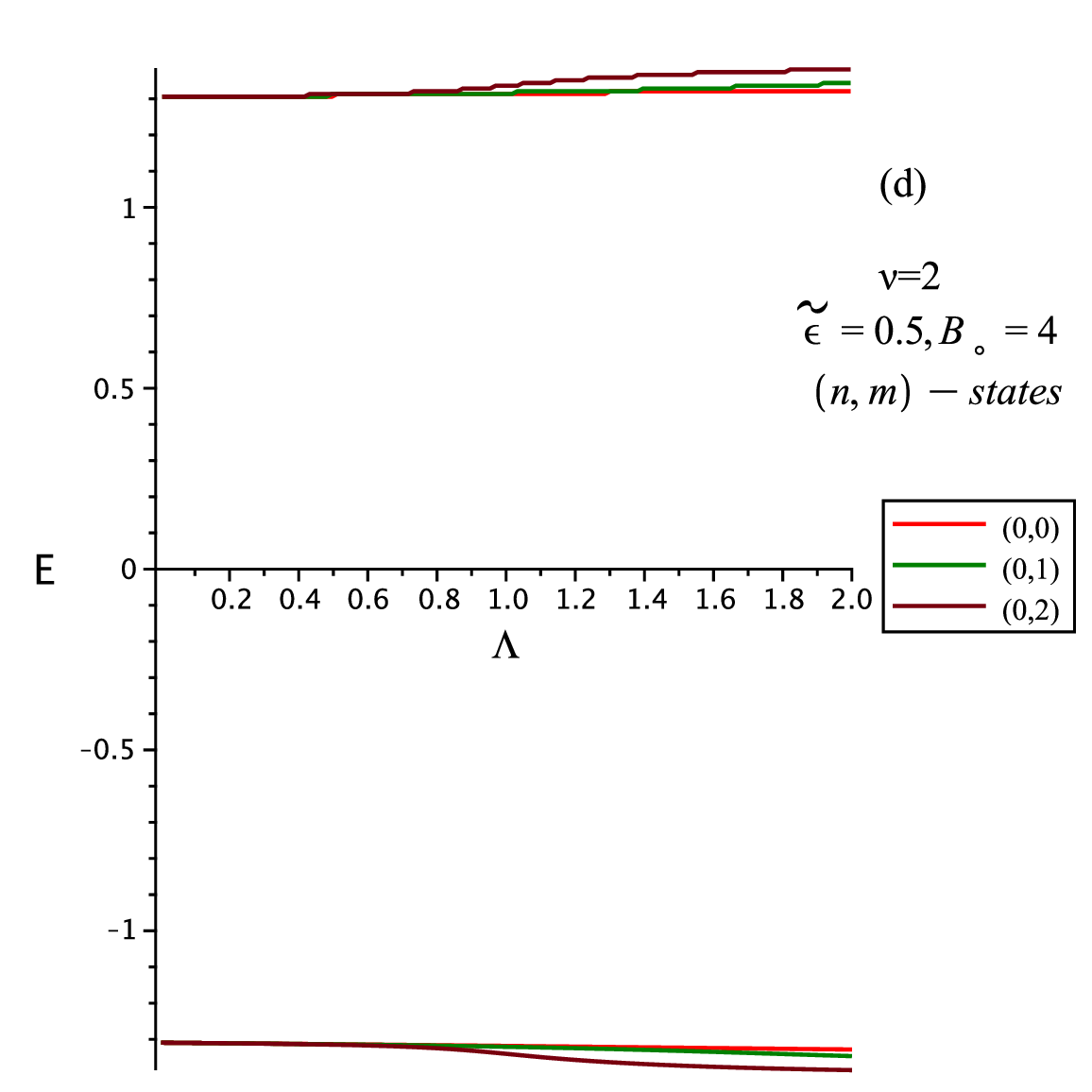}
\caption{\fontsize{7.8}{8.8}\selectfont The figure shows the energy levels for KG particles and antiparticles given by (\ref{III.3.4}), and we plot (a) $E$ against $\tilde{\epsilon}$ for $n=0$, $m=0,1,2,3,4$, $\Lambda=0.5$, and $\mathcal{B}_\circ=1=e$,  (b) $E$ against  $\mathcal{B}_\circ$ for $n=0$ and $m=0,1,2,3,4$, $\Lambda=0.1$, and $\tilde{\epsilon}=0.5$, (c) $E$ against  $\mathcal{B}_\circ$ for $n=1$ and $m=0,1,2,3,4$, $\Lambda=0.1$, and $\tilde{\epsilon}=0.5$, and (d) $E$ against $\Lambda$ for $n=0$ and $m=0,1,2$, $\mathcal{B}_\circ=4$, and $\tilde{\epsilon}=0.5$, (d) $E$ against .}
\label{fig3}
\end{figure*}

\subsection{Rainbow functions pair $f_0(u)=1$, $f_1\left(u \right) = \sqrt{1-\tilde{\epsilon}\left\vert E\right\vert^\nu}$\,; $\nu=1,2$ }

\vspace{0.10cm}
\setlength{\parindent}{0pt}

This pair of rainbow functions is motivated by LQG \cite{R1.38,R1.39}, which has been shown to be fully consistent with the RG principles, thus ensuring invariance of the Planck energy \( E_p \). This energy scale represents the maximum attainable energy for both particles and antiparticles (see, e.g., \cite{R1.34,R1.35,R1.40}). Accordingly, it is of particular interest to explore the behavior of such rainbow function pairs when applied to KG particles and antiparticles in a magnetized BM spacetime. We begin by considering the case \(\upsilon = 1\), which, through (\ref{II.2.9}), leads to the following expression:
\begin{equation}
    E^2 + \tilde{\epsilon}\, \mathcal{G}_{nm} |E| - \tilde{\mathcal{G}}_{nm} = 0. \label{III.3.1}
\end{equation}
The corresponding energy solutions are given by:
\begin{equation}
    E_+ = \frac{-\tilde{\epsilon}\, \mathcal{G}_{nm}}{2} + \frac{1}{2} \sqrt{\tilde{\epsilon}^2 \, \mathcal{G}_{nm}^2 + 4 \tilde{\mathcal{G}}_{nm}}, \label{III.3.2}
\end{equation}
and
\begin{equation}
    E_- = \frac{\tilde{\epsilon} \,\mathcal{G}_{nm}}{2} - \frac{1}{2} \sqrt{\tilde{\epsilon}^2 \,\mathcal{G}_{nm}^2 + 4 \tilde{\mathcal{G}}_{nm}}. \label{III.3.3}
\end{equation}
These energy levels are illustrated in Figure \ref{fig2}. It is evident that, as the cosmological constant \(\Lambda \to \infty\), equation (\ref{III.3.1}) implies that the maximum energy, \( |E|_{\text{max}} \), tends to \( 1/\tilde{\epsilon} = 2 \). This indicates an asymptotic saturation of the energy levels of KG particles and antiparticles in the BM spacetime, as clearly observed in Figure \ref{fig2}. 

Next, we consider the case \(\upsilon = 2\), which yields the following result:
\begin{equation}
    E^2 = \frac{\tilde{\mathcal{G}}_{nm}}{1 + \tilde{\epsilon}\, \mathcal{G}_{nm}} \quad \Rightarrow \quad E_\pm = \pm \sqrt{\frac{\tilde{\mathcal{G}}_{nm}}{1 + \tilde{\epsilon}\, \mathcal{G}_{nm}}}. \label{III.3.4}
\end{equation}
The corresponding energy levels are shown in Figure \ref{fig3}. This plot further confirms the asymptotic behavior of the energy spectrum as the cosmological constant \(\Lambda \to \infty\). As indicated by (\ref{III.3.4}), the maximum energy in this case is given by \( |E|_{\text{max}} \sim \sqrt{1/\tilde{\epsilon}} = \sqrt{2} \approx 1.41 \). This behavior is in full agreement with the analytical expression and is visually represented in Figure \ref{fig3}. Notably, the same behavior observed for the Magueijo-Smolin rainbow functions in Figure \ref{fig1} reemerges in Figures \ref{fig2} and \ref{fig3}. Remarkably, an increase in the uniform magnetic field strength causes all $m$ states to collapse into the $m=0$ state for a given radial quantum number $n$. However, as the cosmological constant $\Lambda$ (of the magnetized BM spacetime) increases, the $m$-related degeneracies are partially lifted, although the degeneracy between states with $m = \pm|m|$ remains preserved as $\Lambda \to 1$. Interestingly, such behavior is absent in the case of scalar KG bosonic particles in a non-uniform magnetic field within BM spacetime, as discussed in \cite{R1.35.1}. The three sets of rainbow functions examined here, a Magueijo-Smolin set and two sets motivated by LQG, are all found to be consistent with the RG-inherited Planck energy invariance.

\begin{figure*}[ht!]
\centering
\includegraphics[width=0.8\textwidth]{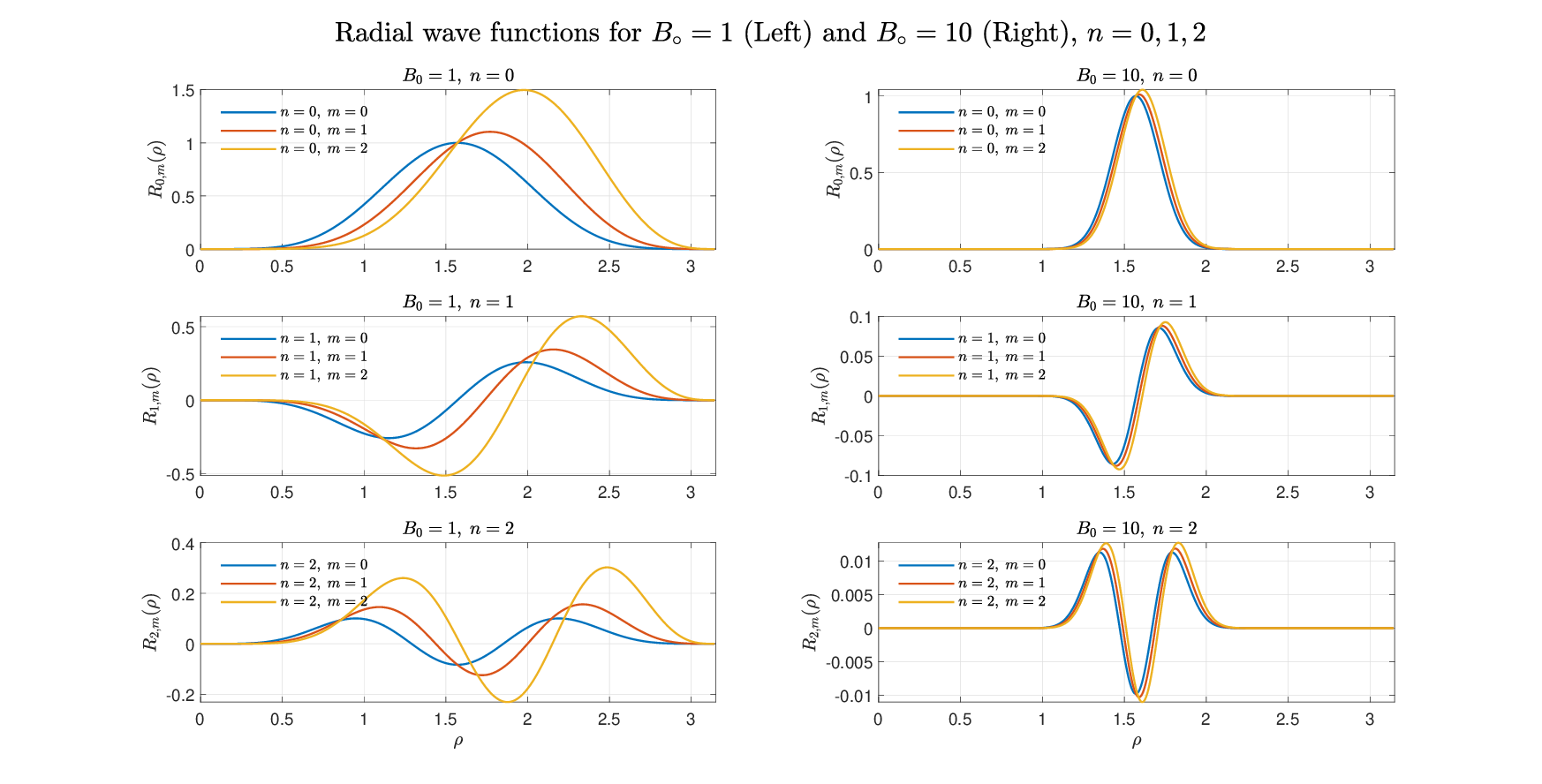}
\caption{\fontsize{7.8}{8.8}\selectfont 
Radial wave functions $R_{n,m}(\rho)$ in (\ref{II.2.10}) for quantum numbers $n = 0,1,2$ and $m = 0,1,2$ under two magnetic field strengths, $\mathcal{B}_\circ = 1$ (left column) and $\mathcal{B}_\circ = 10$ (right column). Each subplot depicts the behavior of $R_{n,m}(\rho)$ over the domain $\rho \in [0, \pi]$, illustrating how the magnetic field intensity affects the spatial distribution of scalar field modes in the magnetized BM universe.}
\label{fig:WFs}
\end{figure*}

\begin{figure*}[ht!]
\centering
\includegraphics[width=0.8\textwidth]{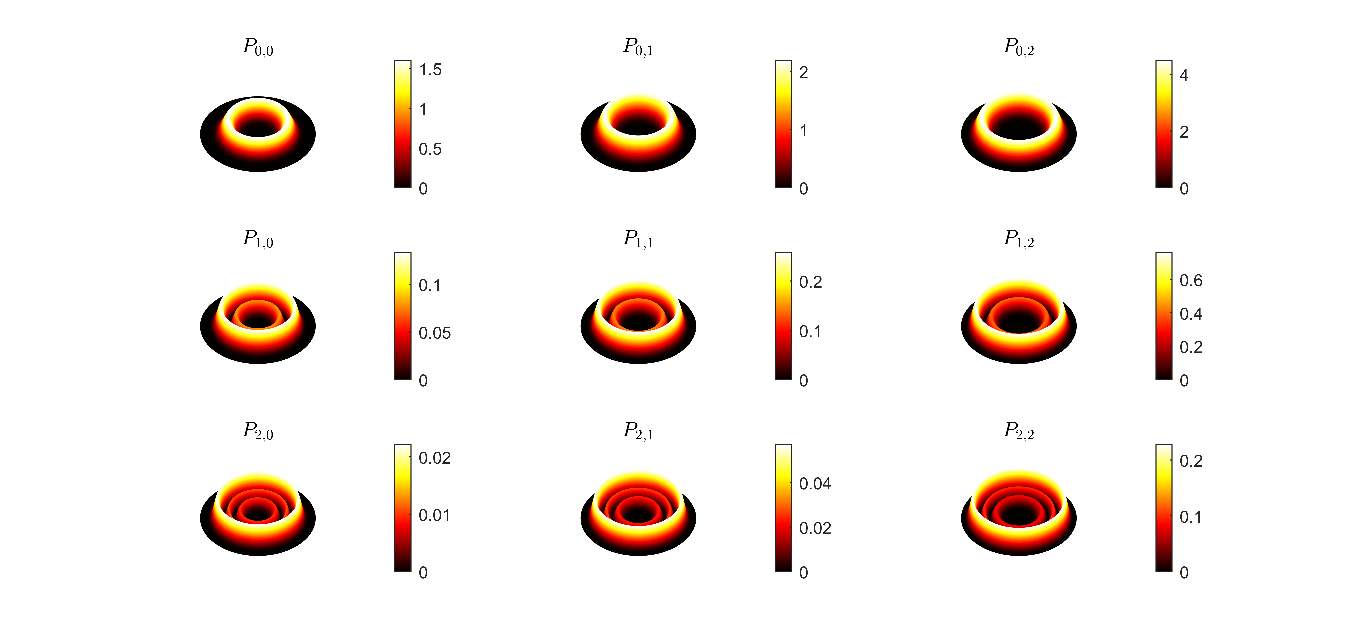}
\caption{\fontsize{7.8}{8.8}\selectfont
Radial probability density distributions \( P_{n,m}(\rho) \) are presented for quantum states with \( n = 0,1,2 \) and \( m = 0,1,2 \), where \(P_{n,m}(\rho) = \int_0^\rho \left| R_{n,m}(\rho') \right|^2 \, \rho' \, d\rho'\). The wavefunctions \( R_{n,m}(\rho) \) are computed under a transverse magnetic field with effective coupling \( \tilde{\mathcal{B}} = \frac{e \mathcal{B}_\circ}{2 \Lambda} \), using \( e = 1 \), \( \mathcal{B}_\circ = 1 \), and \( \Lambda = 0.1 \). The surface plots are mapped to Cartesian coordinates via \( x = \rho \cos\phi \), \( y = \rho \sin\phi \), where both the vertical axis and color scale represent the magnitude of \( P_{n,m}(\rho) \). }
\label{fig:RPDF}
\end{figure*}

\begin{figure*}[ht!]
\centering
\includegraphics[width=0.8\textwidth]{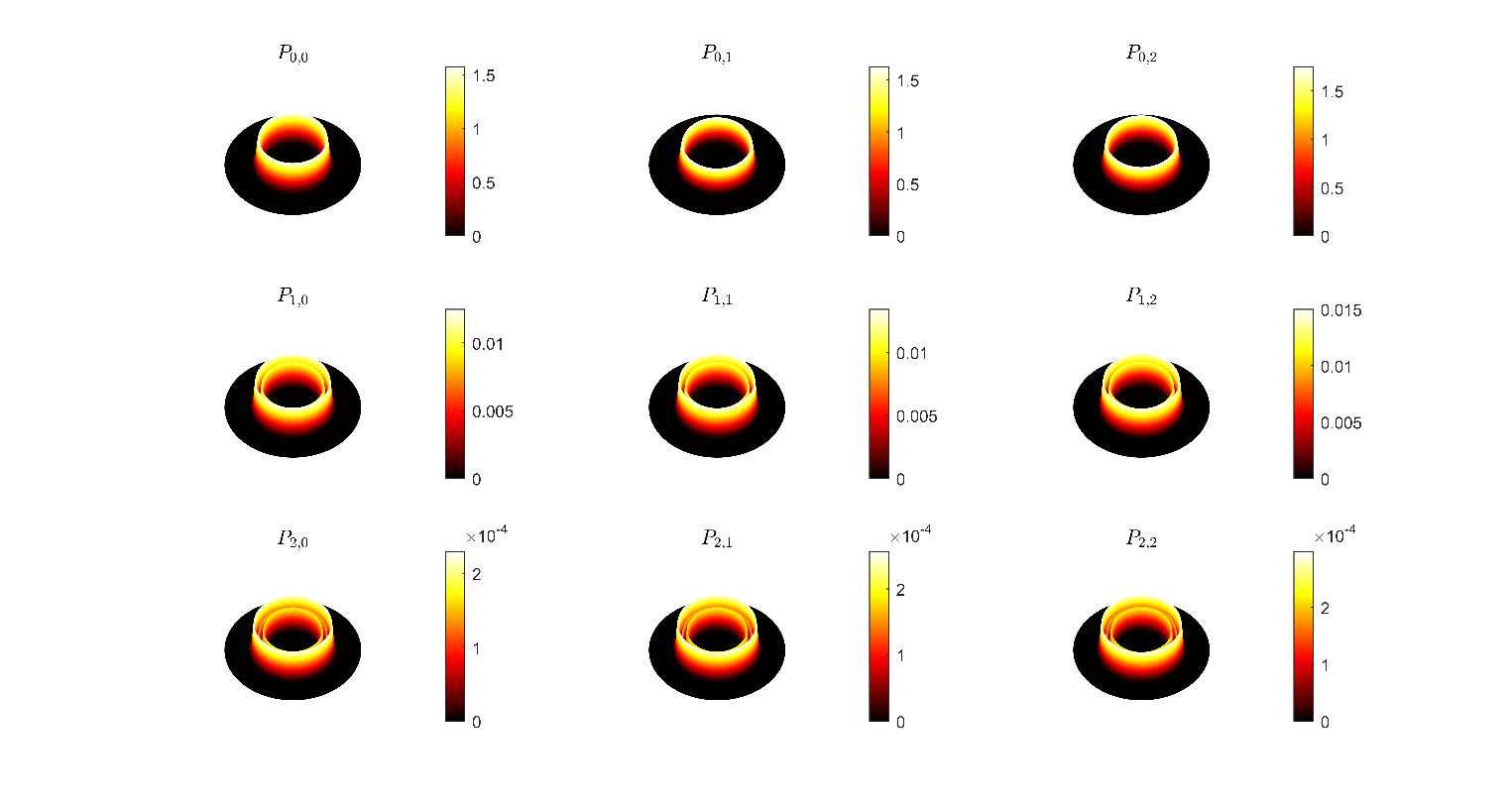}
\caption{\fontsize{7.8}{8.8}\selectfont
Radial probability density distributions \( P_{n,m}(\rho) \) are shown for quantum states labeled by \( (n,m) \), representing the eigenstates of a charged particle confined to a curved surface under a transverse magnetic field. The system parameters are fixed as \( e = 1 \), \( \Lambda = 0.1 \), and \( \mathcal{B}_\circ = 10 \), corresponding to a strong dimensionless magnetic coupling \( \tilde{\mathcal{B}} = \frac{e \mathcal{B}_\circ}{2\Lambda} \gg 1 \). The plots display the spatial distributions of \( P_{n,m}(\rho) \) for discrete states with \( n = 0,1,2 \) and \( m = 0,1,2 \), rendered as three-dimensional surfaces in Cartesian coordinates via the polar mapping \( x = \rho \cos\phi \), \( y = \rho \sin\phi \), with height and colormap indicating the magnitude of the probability density. }
\label{fig:RPDF-2}
\end{figure*}

\section{Concluding Remarks}\label{sec:4}

\vspace{0.10cm}
\setlength{\parindent}{0pt}

In this comprehensive investigation, we have conducted an exact analytical study of KG bosons and antibosons subjected to an external uniform magnetic field in the background of the BM spacetime, modified by the framework of RG and endowed with a positive cosmological constant. This work builds upon recent advances in quantum field theory in curved spacetimes, providing a detailed exploration of how gravitational confinement, topological structures, magnetic interactions, and QG-inspired modifications affect the energy-momentum dispersion relations. A key aspect of our analysis is that the geometry of the BM spacetime naturally induces an infinite sequence of infinite potential barriers, appearing at discrete surfaces determined by the expression $\rho = \sqrt{2\Lambda} r = \tau \pi$, with $\tau = 0, \pi, 2\pi, \dots$. These potential barriers act as impenetrable boundaries, dividing spacetime into an infinite number of confinement regions, each similar to an ideal infinite quantum well with rigid boundaries. Within the first of these confinement domains, we have obtained exact, closed-form solutions to the KG equation, expressed using hypergeometric polynomials for the radial wave functions. This framework allowed us to precisely determine the energy spectrum and examine the effects of external fields and RG corrections.

\vspace{0.10cm}
\setlength{\parindent}{0pt}

Our results reveal several notable features regarding the spectroscopic structure of the system. The inclusion of RG effects, through both Magueijo-Smolin rainbow functions and those inspired by LQG, significantly influences the energy spectrum. In particular, the uniform external magnetic field, which normally lifts the degeneracies associated with the magnetic quantum number $m$, behaves in an unexpected way within this curved spacetime setting. Specifically, the combined effect of the magnetic field and the BM spacetime geometry causes all magnetic quantum states $m$ to collapse onto the $m = 0$ state for a fixed radial quantum number $n$. This behavior differs from what is seen in flat spacetime quantum mechanics and highlights how spacetime curvature and topology can dominate the spectral properties of quantum fields. Furthermore, our analysis shows that the cosmological constant $\Lambda$, which is usually important at large (cosmological) scales, also has a direct effect on the local quantum behavior of the system. When $\Lambda$ increases beyond physically realistic values, a partial lifting of the degeneracies associated with the magnetic quantum number is observed. This indicates a subtle connection between the global geometry of spacetime and the local quantum characteristics of relativistic particles.

\vspace{0.10cm}
\setlength{\parindent}{0pt}

The study also confirms the consistency of the rainbow functions used with the RG principles. In both the Magueijo-Smolin and LQG-inspired cases, the Planck energy $E_p$ appears clearly as the highest possible energy for particles and antiparticles. This maintains the theoretical framework of QG phenomenology and keeps the symmetry of the particle and antiparticle energy spectra around $E = 0$. It is important to note that the unusual degeneracy behavior caused by the uniform magnetic field is not observed in cases with non-uniform magnetic fields in the same BM-RG spacetime \cite{R1.35.1}. This difference emphasizes that both the magnetic field configuration and the geometry of spacetime influence the spectral and confinement properties of quantum fields.

\vspace{0.10cm}
\setlength{\parindent}{0pt}

With reference to the radial wave functions given in equation (\ref{II.2.10}), we plot in Figure \ref{fig:WFs} the radial wave functions \(R_{n,m}(\rho)\) for quantum numbers \(n = 0, 1, 2\) and \(m = 0, 1, 2\) at two magnetic field strengths \(\mathcal{B}_\circ = 1\) and \(10\). As the magnetic field strength increases, the radial wave functions, representing KG particles and antiparticles, become increasingly localized. For \(\mathcal{B}_\circ = 1\), the particles are mainly distributed between \(\rho \sim 0.5\) and \(\rho \sim 3\), while for \(\mathcal{B}_\circ = 10\), they are restricted to the region \(\rho \sim 1\) to \(\rho \sim 2\). This localization effect is further illustrated in the corresponding radial probability densities shown in Figures \ref{fig:RPDF} and \ref{fig:RPDF-2}. Radial probability densities are defined as
\[P_{n,m}(\rho) = \int_0^{\rho} |R_{n,m}(\rho')|^2 \, \rho' \, d\rho',\]
and are visualized in two dimensions via $x = \rho \cos \phi$ and $y = \rho \sin \phi$, where the color intensity represents the magnitude of $P_{n,m}$. The parameter $\tilde{\mathcal{B}} = \frac{e\, \mathcal{B}_\circ}{2\, \Lambda}$ characterizes the effect of the external magnetic field on the quantum states ($\mathcal{B}_\circ=1$ in Figure \ref{fig:RPDF} and $\mathcal{B}_\circ=10$ in Figure \ref{fig:RPDF-2}). As the radial quantum number $n$ increases, the radial probability distributions develop additional nodal rings, corresponding to higher energy states with more radial nodes. For $n=0$, the density is mainly concentrated near the origin, while for $n=1,2$, more complex ring-like structures emerge. For fixed $n$, increasing the azimuthal quantum number $m$ pushes the probability density outward due to the centrifugal barrier, broadening the spatial extent of the wave function. Importantly, increasing the magnetic field strength $\mathcal{B}_\circ$ enhances the magnetic confinement, effectively compressing the probability densities toward smaller radial distances and strengthening localization near the origin. The magnetic field acts as a confining potential that counteracts the outward push caused by larger $m$ values, resulting in sharper and more localized radial peaks. Stronger magnetic fields also slightly shift the nodal positions inward, modifying the spatial probability distribution. These results illustrate how tuning the magnetic field allows control over quantum confinement and the spatial characteristics of the system’s eigenstates. Throughout all states, rotational symmetry is preserved, consistent with conservation of angular momentum. The observed evolution of probability densities with increasing $n$, $m$, and $\mathcal{B}_\circ$ provides a quantitative understanding of how quantum numbers and magnetic field strength shape spatial quantum states under magnetic confinement.

\vspace{0.10cm}
\setlength{\parindent}{0pt}

This work provides an exact and detailed description of the KG boson dynamics in the BM-RG spacetime, showing how gravitational confinement, topological features, magnetic fields, and RG corrections influence the energy spectrum and wave functions. These results contribute to the theoretical understanding of quantum fields in curved and magnetized backgrounds and suggest potential areas for further study. Future work could include extending the analysis to fermionic or higher-spin fields, exploring analogous phenomena in optical or condensed matter systems (in principle), and examining how these effects might appear in astrophysical or cosmological contexts. Although primarily theoretical, the findings help clarify how QG-inspired modifications could affect particle behavior in curved spacetimes.

\vspace{0.10cm}
\setlength{\parindent}{0pt}

Moreover, it should be noted that the BM spacetime considered here is an exact solution of the Einstein-Maxwell equations, describing the self-consistent interaction between a magnetic field and spacetime geometry. With its idealized cylindrical symmetry and infinitely extended magnetic flux tube, it should be regarded primarily as an analytical model rather than a physically realized configuration. Nevertheless, it is useful for capturing several qualitative effects in strongly magnetized, curved environments, such as the geometric confinement of scalar bosons, curvature-dependent energy spectra, and the collapse of magnetic sublevels. Near magnetars, neutron stars, or magnetized accretion disks, similar interactions between strong magnetic fields and spacetime curvature can influence particle localization and energy levels. The inclusion of rainbow gravity corrections further introduces Planck-scale deformations of the dispersion relation, providing a framework to explore high-energy effects. Thus, although not physically observable, the BM spacetime serves as a theoretical laboratory for studying how gravity, magnetism, and rainbow gravity effects jointly influence quantum fields in curved, magnetized backgrounds.










\end{document}